\documentclass[12pt,letterpaper]{article}
\usepackage{graphicx}
\usepackage{amsmath}
\usepackage{amssymb}
\usepackage{hyperref}

\numberwithin{equation}{section}

\usepackage{tikz}
\usetikzlibrary{arrows}
\usetikzlibrary{decorations.text}
\tikzset{>=latex}
\tikzstyle{W}=[draw,circle,scale=1]
\tikzstyle{B}=[draw,circle,fill=black,scale=1]
\tikzstyle{H}=[draw,circle,fill=gray,scale=1]
\tikzstyle{every picture}=[scale=1,baseline=(current bounding box.south)]
\tikzstyle{every loop}=[]

\def\bx{\mathbf{x}}
\def\bq{\mathbf{q}}

\def\cN{\mathcal{N}}
\newcommand{\cO}{\mathcal{O}}
\newcommand{\bZ}{\mathbb{Z}}
\newcommand{\bC}{\mathbb{C}}
\newcommand{\fD}{\mathfrak{D}}
\newcommand{\fA}{\mathfrak{A}}
\newcommand{\CP}{\mathbb{CP}}
\newcommand{\cI}{\mathcal{I}}
\newcommand{\interior}{\lrcorner}
\newcommand{\istar}{*}

\DeclareMathOperator{\vol}{vol}
\newcommand{\tgd}{{\partial}_B{}}
\newcommand{\tgdb}{{\bar{\partial}_B}}
\newcommand{\tgn}{{\nabla}_B{}}
\newcommand{\tgnb}{\bar{\nabla}_B{}}

\DeclareMathOperator{\Tr}{Tr}
\DeclareMathOperator{\SU}{SU}
\DeclareMathOperator{\PE}{PE}
\DeclareMathOperator{\tr}{tr}
\DeclareMathOperator{\ch}{ch}
\DeclareMathOperator{\Todd}{Todd}

\begin{document}

\begin{titlepage}

\begin{flushright}
IPMU 12-0135 \\
UT-12-18\\
RIKEN-MP-72
\end{flushright}
\vskip 2cm

\begin{center}
{\Large \bfseries

\begin{tikzpicture}
\node (A) at (0,0) {};
\node (B) at (10,0) {};
\draw[decorate,decoration={text along path,text align=center,
        text={Superconformal indices,}},out=30,in=150] (A) to (B);
\draw[decorate,decoration={text along path,text align=center,
        text={Sasaki-Einstein manifolds,}},out=0,in=180] (A) to (B);
\draw[decorate,decoration={text along path,text align=center,
        text={and cyclic homologies}},out=-30,in=-150] (A) to (B);
\end{tikzpicture}

}

\vskip 1.2cm

Richard Eager$^\clubsuit$,
Johannes Schmude$^{\clubsuit,\spadesuit}$,
and Yuji Tachikawa$^{\clubsuit,\diamondsuit}$

\bigskip
\bigskip

\begin{tabular}{ll}
$^\clubsuit$ &Kavli Institute for the Physics and Mathematics of the Universe, \\
&University of Tokyo,  Kashiwa, Chiba 277-8583, Japan\\
$^\spadesuit$ &Mathematical Physics Lab., \\
&RIKEN Nishina Center, Saitama 351-0198, Japan \\
$^\diamondsuit$ &Department of Physics, Faculty of Science, \\
&University of Tokyo,  Bunkyo-ku, Tokyo 133-0022, Japan
\end{tabular}

\vskip 1.5cm

\textbf{Abstract}
\end{center}

\medskip
\noindent
The superconformal index of the quiver gauge theory dual to type IIB string theory on the product of an arbitrary smooth Sasaki-Einstein manifold with five-dimensional AdS space is calculated both from the gauge theory and gravity viewpoints.
We find  complete agreement.  Along the way, we find that the index on the gravity side can be expressed in terms of the Kohn-Rossi cohomology of the Sasaki-Einstein manifold and that the index of a quiver gauge theory equals the Euler characteristic of the cyclic homology of the Ginzburg dg algebra associated to the quiver. 

\bigskip
\vfill
\end{titlepage}

\setcounter{tocdepth}{2}
\tableofcontents

\section{Introduction}
The superconformal index (the index) of a four-dimensional
superconformal field theory is the partition function of the theory on
$S^1\times S^3$ with supersymmetric boundary conditions.
Equivalently, the index is the generating function of the number of
operators weighted by their fermion number, so that the contributions
from the long multiplets cancel out
\cite{Romelsberger:2005eg,Kinney:2005ej}:
\begin{equation}
\mathcal{I}(t,y)  = \Tr \;(-1)^F t^{2(E + j_2)} y^{2 j_1}, 
\end{equation} where $E$ is the operator dimension, $F$ is the fermion
number, and $(j_1,j_2)$ are the spins of the operator.

The index is a robust quantity independent of the exact marginal deformations of the theory, and is securely calculable in terms of elliptic beta integrals \cite{Spiridonov:2010em} if one knows the ultraviolet Lagrangian description which flows to the superconformal theory in the infrared, assuming that the superconformal R-symmetry in the infrared can be identified in the ultraviolet \cite{Festuccia:2011ws}. 
This allows us to perform checks of various Seiberg dualities, by calculating the indices using different ultraviolet realizations of the same infrared theory and showing that they agree. This program has been successfully carried out for theories with single gauge groups \cite{Romelsberger:2007ec,Dolan:2008qi,Spiridonov:2009za,Spiridonov:2011hf}. Attempts have also been made to read off other information such as 't Hooft anomaly coefficients from the index \cite{Sudano:2011aa,Spiridonov:2012ww}.

A large class of superconformal field theories is realized as the low-energy limit of the theory on  multiple D3-branes put on the tip of a Calabi-Yau cone in type IIB string theory.  Equivalently, these theories can be described as the holographic duals of type IIB string theory on the product of AdS$_5$ and a Sasaki-Einstein 5-manifold \cite{Klebanov:1998hh}.  Prototypical examples involving orbifolds of $S^5$ and $T^{1,1}$ were studied intensively in the last century. The 2004 discovery of a completely new class of Sasaki-Einstein manifolds \cite{Gauntlett:2004yd,Cvetic:2005ft} reinvigorated the subject.  The corresponding quiver gauge theories were found \cite{Benvenuti:2004dy} and led to the discovery of the field theory duals of all toric Sasaki-Einstein manifolds \cite{Franco:2005sm}.  The algebraic a-maximization procedure for determining the superconformal R-symmetry on the gauge theory side was then mapped to the geometric process of volume minimization on Sasaki manifolds \cite{Martelli:2006yb}.

During these developments, the usefulness of the index to the holographic study of the superconformal theories was not well-appreciated. 
Consequently, the superconformal index was only calculated for the orbifolds of $S^5$ and $T^{1,1}$ \cite{Nakayama:2005mf,Nakayama:2006ur}.   In \cite{Gadde:2010en}, the first significant step was made to study the superconformal index of the `new' Sasaki-Einstein geometries. The authors of \cite{Gadde:2010en} described how to extract the single-trace index\footnote{We do not count the identity operator $1$ as a single-trace operator. } 
\begin{equation}
\mathcal{I}_{s.t.}(t,y)  = \Tr_\text{single trace ops.} \;(-1)^F t^{2(E + j_2)} y^{2 j_1} 
\end{equation}
 from the quiver description of the gauge theory, and observed that the single-trace index has a rather remarkable factorization. They also compared the gauge theory result to the index  calculated from the gravity description, but the results from gravity were available only for the  $S^5$ and $T^{1,1}$ manifolds, based on the classic Kaluza-Klein analysis in \cite{Kim:1985ez} and \cite{Ceresole:1999zs,Ceresole:1999ht}.

The first aim of this paper is to explain how to translate the single-trace index of the gauge theory into a geometric quantity of the Calabi-Yau cone $X$ over the smooth base $Y$: \begin{equation}
ds^2_X = (d\rho)^2+\rho^2 ds^2_Y
\end{equation} where $\rho>0$ is the radial coordinate.\footnote{Note that we do not include the tip $\rho=0$ in the cone $X$ in this paper. This distinction is mathematically relevant, since the cohomology groups on $X$ and on $X\cup \{\text{tip}\}$ can be different. }
 We find that the single-trace index is independent of $y$ and essentially given by \begin{equation}
1+\cI_{s.t.}(t)=\sum_{0\le p-q\le 2} (-1)^{p-q} \Tr  t^{2D} \bigm| H^q(X,\wedge^p \Omega'_X). \label{aaa}
\end{equation}
Here, $X$ is the Calabi-Yau cone, $D$ is the dilatation vector field of the cone, and $\Omega'_X$ is the part of the holomorphic cotangent bundle  $\Omega_X$ perpendicular to $D$.  We write $\Tr A \bigm| V$ for the trace of an operator $A$ acting on a vector space $V$.  By an abuse of notation we, the $D$ in $t^{2D}$ represents the eigenvalue of the Lie derivative along the dilation vector field $D$ acting on the differential forms.  The group $H^q(X,V)$ is the space of harmonic sections of $V\otimes\Omega^{(0,q)}_X$, or equivalently the $q$-th sheaf cohomology valued in a vector bundle $V$. Also, due to various vanishing theorems, the sum is effectively only over $(p,q)=(0,0)$, $(1,0)$, $(2,0)$ and $(1,1)$. 

We will check that the gauge theory index is given by \eqref{aaa} by directly computing \eqref{aaa} and comparing it against the gauge theory formula found in \cite{Gadde:2010en} for general toric Calabi-Yau cones and for cones over del Pezzo surfaces. We will also see that the index of the quiver gauge theory was already introduced under a different name in  2006 in a mathematics paper \cite{Ginzburg:2006fu}.  There, the index was calculated with the same technique yielding the same result as in \cite{Gadde:2010en}. However, the index was called the Euler characteristic of the cyclic homology of Ginzburg's dg algebra associated to the quiver.  Mathematical machinery then allows us (under certain assumptions) to re-express  the Euler characteristic as the expression \eqref{aaa} in general. 

Our second aim is to compare the gauge theory result with the index calculated from the gravity description. For this purpose, we perform the Kaluza-Klein expansion on general Sasaki-Einstein 5-manifolds to identify the shortened multiplets contributing the index. We find that the index on the gravity side is given by \begin{equation}
\sum_{0\le p-q\le 2} (-1)^{p-q} \Tr  t^{2\xi } \bigm| H^{p,q}_{\tgdb}(Y) \label{bbb}
\end{equation} where $Y$ is the Sasaki-Einstein base, $\xi=J \left(r\partial_r\right)$ is the Reeb vector, and $H^{p,q}_{\tgdb}(Y)$ are the Kohn-Rossi cohomology groups of $Y$ under the tangential Cauchy-Riemann differential $\tgdb$, defined as follows. Let $\xi=\cI D$ be the Reeb vector, where $\cI$ is the complex structure of $X$. Let $\eta$ be the corresponding one-form on $Y$. The complexified cotangent bundle of $Y$ can then be decomposed as \begin{equation}
\Omega_Y=\bC \eta \oplus \Omega^{(1,0)}_Y \oplus \Omega^{(0,1)}_Y
\end{equation} where $\Omega^{(1,0)}_Y$ is the holomorphic part under the restriction of $\cI$ on $Y$. 
We  form the bundle $\Omega^{(p,q)}_Y=\wedge^p \Omega^{(1,0)}_Y\otimes \wedge^q\Omega^{(0,1)}_Y $. 
Then $\tgdb$ is the projection of the exterior derivative $d$ which sends a section of $\Omega^{(p,q)}_Y$ to a section of $\Omega^{(p,q+1)}_Y$. The Kohn-Rossi cohomology $H^{p,q}(Y)$ is the cohomology of this complex.

The expressions \eqref{aaa} and \eqref{bbb} agree, because an element in $H^{p,q}_{\tgdb}(Y)$ is always given by restricting an element in $H^q(X,\wedge^p \Omega'_X)$ to $Y$, thus showing that the index calculated from the gauge theory side and the index calculated from the gravity side coincide.

\subsubsection*{Organization}
The rest of the paper is organized as follows.
We first give an overview of our results in section \ref{summary}. 
Then in section \ref{gauge} we compute the the index of the gauge theory. 
In section \ref{supergravity} we study the index by performing the Kaluza-Klein expansion of supergravity modes on a Sasaki-Einstein manifold. 
Finally in section \ref{conclusions} we conclude with a discussion of future directions for research.
We have two appendices.  In appendix~\ref{supergravitydetails} we give details of the supergravity calculations.
In appendix~\ref{abstractmath} we relate the cyclic homology of Ginzburg's dg algebra to the cohomology groups $H^{q}(X,\wedge^p\Omega'_X)$.
In this paper, we assume the Sasaki-Einstein manifold $Y$ is smooth, and the corresponding quiver does not have loops starting and ending at the same node.

\section{Overview of the results}\label{summary}
We first present a summary of our findings.  We follow the conventions of \cite{Gadde:2010en}. The superconformal index is defined as \begin{equation}
\mathcal{I}(t,y,\mu_a)  = \Tr \;(-1)^F t^{2(E + j_2)} y^{2 j_1} \prod_i \mu_a^{F_a},
\end{equation} where the trace is over the Hilbert space of the theory on $S^3,$ or equivalently over the space of operators. Here, $j_{1,2}$ are the left and the right spin, $E$ is the scaling dimension, and $F_a$ are charges under the flavor symmetries; $t$, $y$ and $\mu_a$ are exponentiated chemical potentials, and $F$ is the fermion number. 
Only the short operators satisfying \begin{equation}
 E - 2 j_2 - \frac{3}{2} r =0
\end{equation} contribute to the trace. 

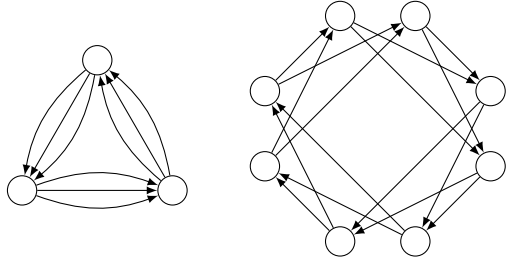
\begin{figure}
\[
\vcenter{\hbox{\begin{tikzpicture}
\node[W] (A) at (0,0) {};
\node[W] (B) at (2,0) {};
\node[W] (C) at (1,1.732) {};
\draw[->] (A) to (B);
\draw[->,out=20,in=160] (A) to (B);
\draw[->,out=-20,in=-160] (A) to (B);
\draw[->] (B) to (C);
\draw[->,out=140,in=280] (B) to (C);
\draw[->,out=100,in=-40] (B) to (C);
\draw[->] (C) to (A);
\draw[->,out=-100,in=40] (C) to (A);
\draw[->,out=-140,in=-280] (C) to (A);
\end{tikzpicture}}}
\qquad\def\A{1.5}\def\B{.5}
\vcenter{\hbox{\begin{tikzpicture}
\node[W] (A1) at (-\A,\B) {};
\node[W] (A2) at (-\A,-\B) {};
\node[W] (C1) at (\A,\B) {};
\node[W] (C2) at (\A,-\B) {};
\node[W] (B1) at (\B,\A) {};
\node[W] (B2) at (-\B,\A) {};
\node[W] (D1) at (\B,-\A) {};
\node[W] (D2) at (-\B,-\A) {};
\draw[->] (A1) to (B1);
\draw[->] (A1) to (B2);
\draw[->] (A2) to (B1);
\draw[->] (A2) to (B2);
\draw[->] (B1) to (C1);
\draw[->] (B1) to (C2);
\draw[->] (B2) to (C1);
\draw[->] (B2) to (C2);
\draw[->] (C1) to (D1);
\draw[->] (C1) to (D2);
\draw[->] (C2) to (D1);
\draw[->] (C2) to (D2);
\draw[->] (D1) to (A1);
\draw[->] (D1) to (A2);
\draw[->] (D2) to (A1);
\draw[->] (D2) to (A2);
\end{tikzpicture}}}
\]

\caption{The quiver for $Y=S^5/\bZ_3$ (left) and for the circle bundle over $dP_5$ (right).\label{fig:quiver-example}}
\end{figure}

Consider a Calabi-Yau cone $X$ over a Sasaki-Einstein 5-manifold $Y$. 
Place $N$ D3-branes at the tip of $X$. 
The low-energy limit of the world-volume theory is a quiver gauge theory consisting of a gauge group $G_v=\SU(k_vN)$
for each vertex $v$,  a chiral field $\Phi_e$ for each edge $e$ which is fundamental under $G_{h(e)}$ and anti-fundamental under $G_{t(e)}$, and a suitable superpotential $W$. 
Here $h(e)$ and $t(e)$ stand for the vertices that are the head and the tail of an edge $e$.
See Fig.~\ref{fig:quiver-example} for the quiver for $Y=S^5/\bZ_3$ and for the circle bundle over $dP_5$.

We are interested in the index $\cI_{s.t.}$ of the single trace operators, \begin{equation}
\mathcal{I}_{s.t.}(t,y,\mu_a)  = \Tr_\text{single trace op.} \;(-1)^F t^{2(E + j_2)} y^{2 j_1} \prod_i \mu_a^{F_a}
\end{equation}
which is related to the full superconformal index in the large $N$ limit by 
\begin{equation}
\mathcal{I}(t,y,\mu_a) \sim \PE[\cI_{s.t.}(t,y,\mu_a)].
\end{equation} Here $\PE$ is the plethystic exponential defined by \begin{equation}
f(t)=\sum_{n\ge 1} a_n t^n \mapsto \PE[f(t)]=\prod_{n\ge 1}\frac{1}{(1-t^n)^{a_n}}. \label{PE}
\end{equation} The plethystic exponential formalizes the relation between multi-trace and single-trace operators.
The full index can be computed by identifying operators contributing the index. This can be done by identifying  the components in a multiplet which give non-zero contributions to the index \cite{Gadde:2010en}. The result  is summarized in Table~\ref{tab:letters}, where $r$ stands for the IR R-charge of the lowest component $\phi$ of a chiral multiplet.
\begin{table}
\centering
\begin{tabular}{|c|c|c|}
\hline
Letter & $(j_1, j_2)$ & $\mathcal{I}$  \\ \hline\hline
$\phi$ & $(0,0)$ & $t^{3r}$ \\ \hline
$\overline{\psi}_2 $& $(0,1/2)$ & $-t^{3(2-r)}$ \\ \hline
\multicolumn{3}{c}{}\\
\hline
\hline
$\partial_{\pm -}$ & $(\pm 1/2, 1/2)$ & $t^3 y^{\pm 1}$ \\
\hline
\end{tabular}
\qquad
\begin{tabular}{|c|c|c|}
\hline
Letter & $(j_1, j_2)$ & $\mathcal{I}$  \\ \hline\hline
$\lambda_1$ & $(1/2,0)$ & $-t^{3} y$ \\ \hline
$\lambda_2 $& $(-1/2, 0)$ & $-t^{3} y^{-1}$ \\ \hline
$\overline{f}_{22}$ & (0,1) & $t^6$ \\ \hline
\hline
$\partial_{\pm -}$ & $(\pm 1/2, 1/2)$ & $t^3 y^{\pm 1}$ \\
\hline
\end{tabular}
\caption{Fields contributing to the index, from a chiral multiplet (left) and from a vector multiplet (right)\label{tab:letters}}
\end{table}

\subsection{Operators associated to holomorphic functions}\label{op-hol}
A fundamental property of the gauge theories we are considering is that the single trace scalar chiral operators consisting solely of the chiral bifundamental fields correspond to the holomorphic functions $f$ on the Calabi-Yau cone $X$. Let us denote by $\cO_f$ a string of chiral bifundamentals such that $\tr \cO_f$ corresponds to $f$. This operator contributes $t^{3r}$ to the single-trace index, where $r$ is the R-charge of $f$.

For each non-constant $f$, we find six short single-trace operators  in the theory, 
consisting of \begin{equation}
\tr \cO_f,\quad
\tr W_\alpha \cO_f,\quad
\tr W_\alpha W^\alpha \cO_f \label{firstNinja}
\end{equation} and \begin{equation}
\tr  \bar W_{\dot\alpha}  \cO_f,\quad
\tr \bar W_{\dot\alpha} W_\alpha  \cO_f,\quad
\tr \bar W_{\dot\alpha} W_\alpha W^\alpha \cO_f.\label{thirdNinja}
\end{equation} 
Here, appropriate insertions of $e^V$ are implied to make the operators gauge invariant, and $W_\alpha$ and $\bar W_\alpha$ are field-strength superfields of a gauge group involved in the string of operators $\cO_f$.  When $\cO_f$ consists of $k$ bifundamentals, there are $k$ choices of field strengths superfields $W_\alpha$ to insert in the trace, but they all give rise to the same element in the chiral ring due to the chiral ring relation $W_{\alpha}^{(h(e))}\Phi_e\sim \Phi_e W_{\alpha}^{(t(e))}$ , and similarly for the other four operators. 
In total, the three operators  \eqref{firstNinja} and their spacetime derivatives contribute \begin{equation}
\frac{t^{3r}}{(1-t^3y)(1-t^3y^{-1})}
-\frac{t^{3r}(t^3 y+t^3 y^{-1}) }{(1-t^3y)(1-t^3y^{-1})}
+\frac{t^{3r} t^3y t^3y^{-1}}{(1-t^3y)(1-t^3y^{-1})}
= t^{3r}\,\phantom{2} \label{firstContribution}
\end{equation} and similarly the three operators \eqref{secondNinja} contribute \begin{equation}
\frac{t^{3(r+2)}}{(1-t^3y)(1-t^3y^{-1})}
-\frac{t^{3(r+2)}(t^3 y+t^3 y^{-1}) }{(1-t^3y)(1-t^3y^{-1})}
+\frac{t^{3(r+2)} t^3y t^3y^{-1}}{(1-t^3y)(1-t^3y^{-1})}
=t^{3(r+2)} 
\label{thirdContribution}
\end{equation} to the single-trace superconformal index,
where $r$ is the R-charge of the holomorphic function $f$. 
Note that the dependence on $y$ disappeared, due to the cancellation of the contributions of the bosonic spacetime derivatives and the fermionic insertions of $W_\alpha$.\footnote{This mechanism has a similar flavor of the roles $W_\alpha$ played in the gauge-theory analysis \cite{Cachazo:2002ry}  of the Dijkgraaf-Vafa matrix model \cite{Dijkgraaf:2002dh}. The authors do not understand the precise relationship between two mechanisms.}

\subsection{Operators associated to holomorphic vector fields}\label{op-vec}
There are also short multiplets of the form 
\begin{equation}
\sum_e c_e\tr  \bar \Phi_e \cO_e \label{holvec}
\end{equation} where $\cO_e$ are strings of chiral bifundamentals, such that the operator \eqref{holvec} is gauge invariant; again the appropriate insertions of $e^V$ is to be understood. 
The operator \eqref{holvec} determines an operation \begin{equation}
\pounds_v: \tr \cO_f \mapsto \tr \cO_g=\sum_e c_e\tr [\cO_e \frac{\partial}{\partial \Phi_e}] \cO_f \label{derivation}
\end{equation} 
Here, $\cO_e \partial/\partial \Phi_e \cO_f$ stands for the operation where we remove a $\Phi_e$ from a string of operators $\cO_f$, and insert $\cO_e$ in its place. 
As an operation from $f$ to $g$, both holomorphic functions on $X$, this is a derivation by a holomorphic vector field $v$ on $X$, so let us denote the operator \eqref{holvec} by $\cO_v$.

We again find three short operators for each holomorphic vector field $v$, given by \begin{equation}
\tr \cO_v, \quad
\tr W_\alpha\cO_v,\quad
\tr W_\alpha W^\alpha \cO_v. \label{secondNinja}
\end{equation} Together, their contribution to the single-trace index is \begin{equation}
-t^{3(r+2)} \label{secondContribution}
\end{equation} where $r$ is the R-charge of the vector field $v$.
Again, we have the cancellation between the insertion of $W_\alpha$ and the spacetime derivatives. 

\subsection{Total single-trace superconformal index}
So far we identified three contributions to the index, \eqref{firstContribution}, \eqref{thirdContribution}, and \eqref{secondContribution}. The contributions in \eqref{firstContribution} are associated to holomorphic functions on the cone $X$. In mathematical terms, the holomorphic functions are elements of $H^0(X,\cO_X)$. 
The contributions in \eqref{secondContribution} are again associated to holomorphic functions on the cone $X$, but with the R-charge shifted by 2.
This shift can be accounted for by multiplying the function by the holomorphic (3,0)-form  on the Calabi-Yau cone, which has the R-charge 2. 
As $X$ is a Calabi-Yau cone, the holomorphic tangent bundle $T_X$ decomposes into $T_X=T'_X \oplus \bC D$ where $D$ is the holomorphic part of the dilatation on the cone. The cotangent bundle is decomposed accordingly; we denote by $\Omega'_X$ the part of $\Omega_X$ perpendicular to $D$.
Then the covariantly-constant $(3,0)$-form is given by $(i\eta+d\rho/\rho)\wedge \Omega$, where $\Omega$ is a standard holomorphic section  $(2,0)$-form of $\wedge^2 \Omega'_X$ of R-charge 2.
Then the operators in \eqref{thirdContribution} are naturally associated to elements in $H^0(X,\wedge^2 \Omega'_X)$ 

Finally, the operators in \eqref{secondContribution} are associated with holomorphic vector fields not involving $D$, i.e. the elements of $H^0(X,T'_X)$. For such a holomorphic vector $v$ of R-charge $r$, $v\interior \Omega$ is an element of $H^0(X,\Omega'_X)$, of R-charge $r+2$, which naturally accounts for the shift by 2 in the exponent in \eqref{secondContribution}. 

As we will see later during our more detailed analysis of the gauge and gravity theories, there is also a contribution from $H^1(X,T'_X)$. This contribution vanishes when $X$ is a toric Calabi-Yau manifold, but is non-zero for non-toric del Pezzo cones. 
If $X$ is compact, a standard result is that its complex structure deformations are elements of $H^1(X,T_X)\simeq H^{1,2}(X).$
The lowest R-charge component of $H^1(X,T'_X)$ indeed gives the complex structure deformations of the del Pezzo base, and induces the deformation of $X$ itself. 

Combining the contributions above, we find that the single-trace index $\cI_{s.t.}$ is given by \begin{equation}
1+\cI_{s.t.}(t,\mu_a)=\sum_{0\le p-q \le 2}(-1)^{p-q} \Tr t^{3r} \prod_i\mu_a ^{F_a} \Bigm|  
H^{q}(X,\wedge^p \Omega'_X) .\label{indexX}
\end{equation} Here $\cO_X=\wedge^0 \Omega'_X$ is the structure sheaf corresponding to the trivial bundle, and we have reinstated the chemical potentials for the mesonic flavor symmetries.
We also used the fact that $H^p(X,\cO_X)=0$ for $p>0$ for a Calabi-Yau cone $X$.

To compare with the supergravity analysis, it is more convenient to phrase the result in terms of the Sasaki-Einstein manifold $Y$, which is the base of the cone $X$. 
On $Y,$ the appropriate notion of the `holomorphy' is given by the so-called tangential Cauchy-Riemann operator $\bar\partial_B$, and the corresponding Kohn-Rossi cohomologies, as explained in the introduction and will be described in more detailed later. 
We then have \begin{equation}
1+\cI_{s.t.}=\sum_{0\le p-q \le 2} \Tr t^{3r} \mu_a^{F_a} \Bigm| H^{p,q}_{\bar\partial_B}(Y). \label{indexY}
\end{equation}

When $Y$ is a regular Sasaki-Einstein manifold, there is a circle fibration  
$\pi: Y \rightarrow S$ over a K\"ahler-Einstein base $S$. The single-trace index in this case can then be phrased purely in terms of the geometry of $S$, which is \begin{equation}
1+\cI_{s.t.}=
\sum_n  t^{2n} \sum_{0\le p-q\le 2} (-1)^{p-q} \Tr  \mu_a^{F_a} \Bigm| H^q(S, (-K_S)^{\otimes n}\otimes\Omega^{(p,0)}_S).
\label{indexS}
\end{equation}
This follows immediately from \eqref{indexX}, because the a section of the bundle $\Omega'_X$ on $X$ with R-charge $2n/3$  comes from a section $\Omega_S \otimes (-K_S)^{\otimes n}$ on $S$.

In section \ref{gauge}, we will see how the study of the gauge theory leads to the expression \eqref{indexX},
and in section \ref{supergravity}, we will see how the Kaluza-Klein expansion on the Sasaki-Einstein manifold give rise to the sum \eqref{indexY}. 

\subsection{Further simplification of the index}

Our formulae \eqref{indexX}, \eqref{indexY}, \eqref{indexS} for the single-trace index phrased in terms of the geometry of the Calabi-Yau cone $X$, the Sasaki-Einstein manifold $Y$, or the K\"ahler-Einstein base $S$ if available, are already quite aesthetically pleasing, but can in fact be  further simplified. We now show how this simplification arises. 

\subsubsection{Toric Calabi-Yaus}\label{sec:toricsimplification}
Let us consider a toric Calabi-Yau cone $X$. 
In this case, there is one superconformal R-symmetry and two mesonic flavor symmetries. We will take a new basis of these symmetries such that the exponentiated chemical potentials are given by $x_1,x_2,x_3$ with $t^6=x_1x_2x_3$.
Then each holomorphic function $f$ has integer charges $\bq=(q_1,q_2,q_3)$ under the three isometries, and contributes  
$\bx^\bq=x_1{}^{q_1}x_2{}^{q_2} x_3{}^{q_3}$  to the index. It is well known that the charges form a cone $M \subset \bZ^3$ and \begin{equation}
\Tr \bx^\bq \bigm|H^0(X,\cO_X)= \sum_{\bq\in M} \bx^\bq .
\end{equation} We easily have \begin{equation}
\Tr \bx^\bq \bigm|H^0(X,\wedge^2 \Omega'_X)= \sum_{\bq\in M} \bx^{\bq+(1,1,1)}.
\end{equation}
The groups $H^{\ge 1}(X,\Omega_X) $ vanish.  The elements of
$H^0(X,\Omega'_X)$ are found using  \begin{equation}
\Tr \bx^\bq \bigm| H^0(X,\Omega'_X)
=\sum_{\bq\in M} n_{\bq} \bx^\bq,
\end{equation} where \begin{equation}
n_q=\begin{cases}
0 & \text{if $q$ is on a 1-dimensional edge  of $M$},\\
1 & \text{if $q$ is on a 2-dimensional face of $M$},\\
2 & \text{if $q$ is on the bulk of $M$}.
\end{cases}
\end{equation}
These theorems are standard in toric geometry and are nicely explained in \cite{Cox}, see in particular Proposition 8.2.18 of \cite{Cox} and Theorem 4.3 in \cite{MR495499}. 
Then it is clear that \begin{align}
1+\cI_{s.t.}&=\Tr \bx^\bq |H^0(X,\cO_X)- \Tr \bx^\bq |H^0(X,\Omega'_X)+ \Tr \bx^\bq |H^0(X,\wedge^2 \Omega'_X) \\
&=1+\sum_{\bq\;\text{on an edge of $M$}} \!\!\! \bx^\bq 
=1+\sum_{i} \frac{\bx^{\bq_{(i)}}}{1-\bx^{\bq_{(i)}}}, \label{toricfac}
\end{align} where $\bq_{(i)}=(q_{(i),1},q_{(i),2},q_{(i),3})$ is the lattice point on the $i$-th edge closest to the origin. 
This explains the observation of \cite{Gadde:2010en} that the single-trace index is just a sum of contributions from the extremal chiral operators identified in \cite{Benvenuti:2005cz}.

\subsubsection{Cones over del Pezzo surfaces}
Next, let us suppose that our Calabi-Yau cone $X$ is a complex cone over the del Pezzo surface $dP_k$ of degree $9 -k$, which is  obtained by blowing up $k$ points on $\CP^2$. We assume $4\le k \le 9$, so that the cone is non-toric. 
Our expression in equation \eqref{indexS} can be  succinctly written as 
\begin{equation}
\cI_{s.t.}= \chi(S, V),
\end{equation}
where  $V$ is the virtual vector bundle \begin{equation}
V =  \sum_{n \ge 1} t^{2n}  (-K_{S})^{\otimes n} \otimes \mathcal{E}, \qquad
\mathcal{E} =  \oplus (-1)^k \wedge^k \Omega_S,
\end{equation}  and $\chi(S,V)$  is the holomorphic Euler characteristic \begin{equation}
\chi(S,V)=\sum_q (-1)^q \dim H^q(S,V).\label{qfoo}
\end{equation}
We have \begin{equation}
\ch(\mathcal{E}) = \ch( \oplus (-1)^k \wedge^k \Omega_S) = (-1)^{\dim S} c_\text{top},
\end{equation}
where $c_\text{top}$ is the top-degree Chern class of $S$.
We then use the Hirzebruch-Riemann-Roch theorem to compute \begin{equation}
\chi(S, (-K_S)^{\otimes n}\otimes \mathcal{E}) = \int_S e^{nc_1} \ch(\mathcal{E}) \Todd(S) = \int_S c_{top}(S) = k+3,
\end{equation} which is independent of $n$.
It then follows that the single-particle index equals
\begin{equation}
\cI_{s.t.}= (k+3) \sum_{n\ge 1}  t^{3n} = (k+3) \frac{t^3}{1-t^{3}}.\label{dPindex}
\end{equation}
Therefore, the single trace superconformal index behaves as if there are $k+3$ `edges' as in \eqref{toricfac}, each with the sequence of R-charges $2,4,6,$ \ldots.  Note that we have this simple result because we sum over $q$ in \eqref{qfoo}. In particular, $H^1(S,\Omega_S\otimes(-K_S)) \simeq H^1(S,T_S)$ is $2k-8$ dimensional, corresponding to the complex structure deformations of $S$.

\section{Gauge theory}\label{gauge}
\subsection{Review of the large $N$ evaluation of the index}
Let us  briefly review the computation of the index of a superconformal field theory with a weakly-coupled UV Lagrangian description \cite{Romelsberger:2007ec,Gadde:2010en}.  
For simplicity, we set the exponentiated chemical potentials of flavor symmetries to be 1, unless otherwise noted. They can be easily reinstated. We assume that the quiver does not contain loops starting and ending at the same node. 

For a quiver gauge theory with chiral multiplets labeled by edges $e \in E$ and gauge fields represented by vertices $v \in V$ we can
define the single-letter index
\begin{equation}
i(t,y; U_v) = \sum_{e \in E} i_{\chi(r)}(t,y;U_{h(e)},U_{t(e)}) + \sum_{v \in V} i_V(t,y;U) \label{single}
\end{equation}
as the sum over all the fundamental fields (``letters'') contributing to the trace. Here $U_v$ is the exponentiated chemical potential for the gauge group $\SU(k_v N)$ associated to the vertex $v$.
These letters must satisfy
\begin{equation}
 E - 2 j_2 - \frac{3}{2} r =0.
\end{equation} 
The single-letter index of a chiral multiplet with R-charge $r$ is
\begin{equation}
i_{\chi(r)}(t,y;U) = 
i_{\phi(r)}(t,y;U)+i_{\bar\psi(r)}(t,y;U),
\end{equation}
where
\begin{equation}
  \begin{aligned}
    i_{\phi(r)}(t,y;U) &= \frac{t^{3r} \chi_R(U)}{(1 - t^3 y)(1-t^3 y^{-1})}, \\
    i_{\bar\psi(r)}(t,y;U) &= -\frac{t^{3(2-r)} \chi_{\overline{R}}(U)}{(1 - t^3 y)(1-t^3 y^{-1})}.    
  \end{aligned}
\end{equation}
Similarly the single-letter index of a vector multiplet is
\begin{equation}
i_V(t,y; U) = \frac{2t^6 - t^3(y + \frac{1}{y})}{(1 - t^3 y)(1-t^3 y^{-1})} \chi_\text{adj}(U).
\end{equation}
These formulas can be reproduced using the table of contributing letters, see Table~\ref{tab:letters}.

Now, the index of the gauge theory is obtained by first taking the plethystic exponential \eqref{PE} of the single letter index \eqref{single} to enumerate  arbitrary words constructed from the single letters, and then projecting to the gauge-invariant operators by integrating over $U$: 
\begin{equation}
\cI(t,y) = \int \prod_v[dU_v] \PE[i(t,y;U_v)].
\end{equation}
In the large $N$ limit  the matrix integral is evaluated using the saddle-point method \cite{Kinney:2005ej,Gadde:2010en}.  The result is that the superconformal index for $\SU(N)$ gauge group is 
\begin{equation}
\mathcal{I}(x) = \prod_{k} \frac{e^{-\frac{1}{k} \Tr i(x^k)}}{\det(1 - i(x^k))}.
\end{equation} Here, $i(x)\equiv i(t,y)$ is a matrix of size $n_v\times n_v,$ where $n_v$ is the number of vertices of the quiver, and is
given by \begin{equation}
i(t,y)= \sum_v  i_V(t,y) E_{v,v} + \sum_e  i_{\phi(r)}(t,y) E_{h(e),t(e)} +\sum_e  i_{\bar\psi(r)}(t,y) E_{t(e),h(e)},
\end{equation}
where $E_{v,w}$ is a matrix such that the $(v,w)$ entry is 1 and all other entries are zero. For example, for the quiver in the left hand side of Fig.~\ref{fig:quiver-example}, we have \begin{equation}
i(t,y)=\left(\begin{array}{rrr}
i_V(t,y) & 3 i_{\phi(2/3)}(t,y) & 3 i_{\bar\psi(2/3)}(t,y)  \\
3 i_{\bar\psi(2/3)}(t,y) & i_V(t,y) & 3 i_{\phi(2/3)}(t,y)  \\
 3 i_{\phi(2/3)}(t,y)  & 3 i_{\bar\psi(2/3)}(t,y) & i_V(t,y)  
\end{array}\right).
\end{equation}
Finally, the single-trace superconformal index can be extracted from the multi-trace index using the plethystic logarithm
\begin{equation}
\mathcal{I}_{s.t.}  = \sum_{n = 1}^{\infty} \frac{\mu(n)}{n} \log \mathcal{I}(x^n) 
 = - \sum_{n = 1}^{\infty} \frac{\varphi(n)}{n} \log [ \det(1 - i(x^n))] - \Tr i(x). \label{PL1}
\end{equation}

\subsection{Further manipulation}
We see that the main quantity entering the expression for the large-$N$ superconformal index is the determinant
$\det(1 - i(t,y)).$
Since $i_V$ is on the diagonal, we can use
\begin{equation}
1 - i_V(t,y) = 1 -  \frac{2t^6 - t^3(y + \frac{1}{y})}{(1 - t^3 y)(1-t^3 y^{-1})} =  \frac{1 - t^6}{(1 - t^3 y)(1-t^3 y^{-1})}.
\end{equation}
By clearing common denominators, we find that \begin{equation}
1 - i(t,y) = \frac{\chi(t)}{(1 - t^3 y)(1-t^3 y^{-1})}, \label{bosh}
\end{equation} where $\chi(t)$ is independent of $y$ and is given by
\begin{equation}
\chi(t) = 1 - M_Q(t) + t^6 M^{T}_Q(t^{-1}) - t^6.\label{chi}
\end{equation}
Here $1$ is the identity matrix and $M_Q(t)$ is the weighted adjacency matrix 
\begin{equation}
M_Q(t) = \sum_{e} t^{3R(e)} E_{h(e),t(e)},
\end{equation}
where $R(e)$ is the r-charge of the edge $e$. 
For example, for the quiver in the left hand side of Fig.~\ref{fig:quiver-example}, we have \begin{equation}
M_Q(t)=\begin{pmatrix}
0 & 3 t^{2} & 0 \\
0 & 0 & 3 t^{2}  \\
3t^2 & 0 & 0 
\end{pmatrix}.
\end{equation}
If there are no adjoint chiral fields then \eqref{PL1} becomes \begin{equation}
\mathcal{I}_{s.t.}(t,y)= - \sum_{n = 1}^{\infty} \frac{\varphi(n)}{n} \log [ \det(\chi(t^n))]  \label{PL2}
\end{equation} due to the cancellation from the denominator of \eqref{bosh} and the subtraction of the trace in \eqref{PL1}.
Note that this expression is now independent of $y$.  If the geometry has a gauge theory description with adjoint chiral fields, such as $\bC^3$ or non-isolated singularities, we can easily account for their contribution to the trace, but our expressions become $y$-dependent.

Let us now reinstate the chemical potentials $\mu_a$ of flavor symmetries. 
Suppose furthermore that the determinant of $\chi(t,\mu_a)$ factorizes \begin{equation}
\det(\chi(t,\mu_a))= \prod_{i=1}^{n_v} (1-t^{r_i}\prod_a\mu_a{}^{f_{i,a}}).
\end{equation} 
Then the plethystic logarithm \eqref{PL2} can be easily evaluated, and gives \begin{equation}
\cI_{s.t.}(t,\mu_a)=\sum_i \frac{t^{r_i}\prod_a\mu_a{}^{f_{i,a}}}{1-t^{r_i}\prod_a\mu_a{}^{f_{i,a}}} .
\end{equation}  

For example, for the quiver for $dP_{k\ge 4}$  given in \cite{Wijnholt:2002qz,Wijnholt:2005mp} and shown in Fig.~\ref{fig:quiver-example} for $k=5$, we find $\det(\chi(t))=(1-t^3)^{k+3}$ via explicit calculations, and obtain the single-trace index \begin{equation}
\cI_{s.t.}(t)=(k+3)\frac{t^3}{1-t^3},
\end{equation} which agrees with the index obtained geometrically in \eqref{dPindex}.

\subsection{Factorization for toric Calabi-Yaus}
In the last step of the calculation in the previous subsection, we assumed the determinant of the matrix $\chi(t)$ factorizes.
We now prove that the determinant of $\chi(t)$ factorizes for a toric Calabi-Yau cone. As in section \ref{sec:toricsimplification}, we re-introduce three $U(1)$ charges so that the exponentiated chemical potentials satisfy $t^2=x_1x_2x_3$, and denote $(x_1,x_2,x_3)$ collectively as $\bx$.
Now, recall that the Hilbert series \begin{equation}
h(\bx)\equiv \Tr\bx^\bq | H^0(X)=\sum_{\bq \in M} \bx^\bq
\end{equation}
of the Calabi-Yau $X$ equals the $(i,i)$ component of the inverse of $\chi(\bx)$,
\begin{equation}
\chi(\bx)^{-1} = \frac{C(\bx)}{\det{\chi(\bx)}} \label{pppp}
\end{equation}
for a suitable choice of the vertex $i$ \cite{MR2355031, Ginzburg:2006fu, Eager:2010yu}.  
Here $C(\bx)$ is the cofactor matrix of $\chi(\bx)$, and is polynomial in $bx$.
Now, it is proven  in Theorem 4.6.11 of \cite{MR847717} that the Hilbert series has the expression as an irreducible fraction \begin{equation}
h(\bx)=\frac{P(\bx)}{D(\bx)} \quad \text{where}\quad  D(\bx)={\prod_{\bq \in CF(M)} (1 - \mathbf{x}^{\mathbf{q}})},\label{qqqq}
\end{equation} where $CF(M)$ is the set of lattice points in $M$ that are not positive-integral linear combinations of other lattice points in $M$.
Comparing \eqref{pppp} and \eqref{qqqq}, we find that $D(\bx)$ divides $\det(\chi(\bx))$. 

We now show that the polynomials $D$ and $\det \chi$ have the same degree so they must in fact be equal.
As $\chi(\bx)$ has degree $2$ in $t$, the determinant has degree $2n_v$. It is known that the number of the gauge groups $n_v$ is equal to the twice of the area of the toric diagram. 
Suppose the toric diagram has vertices $v_1,v_2, \dots, v_k$, all on a plane defined by $w$, such that $\langle w, v_i \rangle=1$ for all $i$.  This condition ensures that we have a toric Calabi-Yau cone of dimension 3.  For simplicity we further assume that the vertices are cyclically ordered and that there are no interior lattice points on the segment connecting two cyclically adjacent vertices $v_i$ and $v_{i+1}$.  This condition ensures the smoothness of the geometry away from the tip of the cone.

The charges of the holomorphic functions are integral lattice points $\mathbf{x}$ defined by the conditions $\langle v_i , \mathbf{x} \rangle \ge 0$ for all $i$. So, the edges of the cone of charges are given by the condition $\langle v_i , z \rangle = \langle v_{i+1}, \mathbf{x} \rangle =0$. One such $\mathbf{x}$ is given by $\bq_i = v_i \times v_{i+1}$, where $\times$ is the cross product. The requirement that there are no integral points on the segment between $v_i$ and $v_{i+1}$ is equivalent to the fact that this $\mathbf{x}$ is the closest lattice point on this edge. This $\bq_i$ clearly belongs to $CF(M)$.
Its degree of $t$ is then $(2/3) \langle w, (v_i \times v_{i+1})\rangle $. 
Therefore, the sum of the degrees of $t$ of $\bx^{\bq_i}$ is \begin{equation}
	\frac23\sum_i \langle w , (v_i \times v_{i+1}) \rangle ,
\end{equation} which is four times the area of the toric diagram and equals $2n_v$. This means that the $\bq_i$ exhausts $CF(M)$, and $D(\bx)=\det(\chi(\bx))$.

\subsection{Superconformal index and  Ginzburg's DG algebra}\label{ginzburg}

So far, we saw that one reason for the simplification of the index is the cancellation between the contribution from the insertions of $W_\alpha$ and the insertions of the spacetime derivatives, in the case of quiver gauge theories.
Once this is taken into account\footnote{The cancellation works except for the single-trace operators without  any letter from chiral multiplets, e.g.  the would-be triple of operators $\tr \bar W_{\dot\alpha}$, $\tr \bar W_{\dot\alpha} W_\alpha$, $\tr \bar W_{\dot\alpha} W^\alpha W_\alpha$. Among these three, the first one is zero because we consider $\SU$ gauge groups, thus spoiling the cancellation of $y$-dependent terms. This can introduce a difference between the superconformal index and the Euler characteristic of the cyclic homology of a term of the form $at^6$ for an integer $a$. We use $\doteq$ in \eqref{sti} and \eqref{stii} to signify this possible discrepancy.  Explicit calculation suggests that it is always just $t^6$. Assuming this,  the formula \eqref{geometric-formula} holds literally. }, the superconformal index gets contributions from the letters $\phi$, $\bar\psi_2$ from the chiral multiplet in the bifundamental, and $\bar f_{22}$ in the vector multiplet, as listed in Table~\ref{tab:letters2}. Among them, the supersymmetry transformation $\delta$ used to define the superconformal index acts as \begin{align}
\delta\phi_e & = 0 , \label{x1}\\
\delta\bar\psi_{e,2} & = \partial W(\phi_e)/\partial \phi_e, \label{x2}\\
\delta\bar f_{v,22} & = \sum_{h(e)=v} \phi_e \bar \psi_{e,2} - 
\sum_{t(e)=v}  \bar \psi_{e,2} \phi_e.\label{x3}
\end{align} 
We can then assign charges $F=0$, $F=1$ and $F=2$ to $\phi,$ $\bar\psi_2$, and $\bar f_{22}$ respectively.  The charge is twice the spin $j_2$ and the  transformation $\delta$ decreases this charge by one.  Calculating the single-trace superconformal index then reduces to evaluating \begin{equation}
\cI_{s.t.}(t)\doteq\Tr (-1)^F t^{3R} | (\text{cyclic gauge invariants made of $\phi_e$, $\bar\psi_{e,2}$ and $\bar f_{v,22}$}).\label{sti}
\end{equation}

\begin{table}
\centering
\begin{tabular}{|c|c|c|}
\hline
Letter & $(j_1, j_2)$ & $\mathcal{I}$  \\ \hline\hline
$\phi$ & $(0,0)$ & $t^{3r}$ \\ \hline
$\overline{\psi}_2 $& $(0,1/2)$ & $-t^{3(2-r)}$ \\ \hline
\end{tabular}
\qquad
\begin{tabular}{|c|c|c|}
\hline
Letter & $(j_1, j_2)$ & $\mathcal{I}$  \\ \hline\hline
$\overline{f}_{22}$ & (0,1) & $t^6$ \\ \hline
\multicolumn{3}{c}{}
\end{tabular}
\caption{Fields contributing to the index, from a chiral multiplet (left) and from a vector multiplet (right), after the cancellation of $W_\alpha$ and the spacetime derivatives $\partial_\mu$ are taken into account.\label{tab:letters2}}
\end{table}

Remarkably, Ginzburg introduced exactly the same fields $\phi_e$, $\bar\psi_{e,2}$ and $\bar f_{v,22}$ and the differential $\delta$ for a quiver $Q$ with a superpotential $W$ in 2006 in \cite{Ginzburg:2006fu}. There, the fields are denoted by $x_e$, $x^*_e$ and $t_v.$
Let us consider a modified quiver $\hat Q$, whose edges consist of all the edges of $Q$, together with   a reverse edge $\tilde e$ for each $e$ and a loop edge $\tilde v$ at each vertex $v$.
Associate variables $X_E$ for each edge $E$ of $\hat Q$ so that $X_e=x_e$, $X_{\tilde e}=x^*_e$ and $X_{\tilde v}=t_v$. 
Ginzburg's differential-graded (DG) algebra $\fD$ is then generated by non-commutative elements $X_E$ with the relation $X_E X_{E'}=0$  unless $t(E)=h(E')$, with the action of the derivation $\delta$ given by \eqref{x1}, \eqref{x2} and \eqref{x3}.
Note that any basis monomial in $\fD$ is given by choosing a (possibly open) path on $\hat Q$, and multiplying $X_E$ for edges on the path. 

Let $[\fD,\fD]$ be a $\bC$-linear space spanned by the commutators of two elements in $\fD$. Then, it is easy to see that the basis of $\fD_\text{cyc}=\fD/(\bC+[\fD,\fD])$ corresponds to the set of closed path of $\hat Q$, or equivalently, the single-trace operators formed from $\phi_e$, $\bar\psi_{e,2}$ and $\bar f_{v,22}$. 
We would like to consider the single-trace operators, up to the pairing given by the supersymmetry transformation $\delta$. This corresponds to taking the homology $H_*(\fD_\text{cyc},\delta)$ with respect to the action of $\delta$ on $\fD_\text{cyc}$. This homology is known as the reduced cyclic homology\footnote{The relevance of the cyclic homology to the quiver gauge theory was first pointed out and developed in e.g.~\cite{Berenstein:2000ux,Berenstein:2000te,Berenstein:2001jr,Berenstein:2002fi}. } of the algebra $\fD$, and is usually denoted by $\overline{HC}_*(\fD)$.
Therefore, the single-trace index \eqref{sti} is now \begin{equation}
\cI_{s.t.}(t)\doteq\Tr (-1)^F t^{3R} | \fD_\text{cyc}= \sum_{i} (-1)^i \Tr t^{3R} | \overline{HC}_i(\fD).\label{stii}
\end{equation}

So far, we have only formally rewritten the single-trace index and have not gained any new insight.
The single-trace index was already evaluating in \cite{MR2330156, MR1191088} and simplified to the form \eqref{PL2}, using essentially the same method independently rediscovered in \cite{Gadde:2010en}. The advantage of reformulating the gauge theory index in terms of cyclic homology is that the cyclic homology groups can be directly related to the supergravity index.
As will be explained in more detail in the Appendix~\ref{abstractmath}, 
 we have \begin{align}
\bC\oplus \overline{HC}_0(\fD) &=   H^0(\wedge^0 \Omega'_X)\oplus H^1(\wedge^1 \Omega'_X)\oplus H^2(\wedge^2 \Omega'_X), \\
\overline{HC}_1(\fD) &=   H^0( \wedge^1\Omega'_X)\oplus H^1(\wedge^2 \Omega'_X),\\
\bC\oplus \overline{HC}_2(\fD) &= H^0(\wedge^2 \Omega'_X),
\end{align}  assuming a few mathematical results which are explained in the appendix.
We conclude that the single-trace index is \begin{equation}
1+\cI_{s.t.}(t)=\sum_{0\le p-q\le 2} (-1)^{p-q} \Tr t^{3R} | H^q(\wedge^p \Omega'_X).\label{geometric-formula}
\end{equation}

\section{Supergravity}\label{supergravity}
In this section, we perform the Kaluza-Klein expansion of type IIB supergravity on AdS$_5$ times a five-dimensional Sasaki-Einstein manifold, and identify the structure of the superconformal multiplets. This analysis was originally done for $S^5$ in \cite{Kim:1985ez}\footnote{Also see a recent review \cite{vanNieuwenhuizen:2012zk}.} and for $T^{1,1}$ in \cite{Ceresole:1999zs,Ceresole:1999ht}.  In those papers, the fact that these manifolds are homogeneous is used to its full extent in order to determine the complete spectrum of the Kaluza-Klein fields. On a general Sasaki-Einstein manifold, the determination of the complete spectrum is too much to be desired, but as we will see below, we can still identify the structure of all the superconformal multiplets. 
In this paper, we will only consider the bosonic components of the multiplets. 

\subsection{Expansion on general Einstein manifolds}
\begin{table}
\[
\begin{array}{l|l|l}
\text{AdS$_5$ mode} & \text{Mass eigenvalue} & \text{10d origin}\\
\hline
H_{(\mu\nu)} & \Delta_0 & g_{\mu\nu}  \\
\hline
B_\mu & \Delta_1+4 + 4\sqrt{\Delta_1+1}  & g_{\mu a} + C_{\mu abc} \\
\phi_\mu & \Delta_1+4 - 4\sqrt{\Delta_1+1} & g_{\mu a} + C_{\mu abc}\\
a_\mu, a_\mu^* & \Delta_1 & C_{\mu a} \\
\hline 
\pi & \Delta_0 + 16 + 8 \sqrt{\Delta_0+8}  & g_{aa} + C_{abcd}  \\
b & \Delta_0+16 -8\sqrt{\Delta_0+8} & g_{aa} + C_{abcd}\\
B,B^*& \Delta_0 &  \imath e^{\varphi}+ C  \\
\phi & \Delta_L-8  & g_{(ab)}  \\
a,a^*& Q^2 + 4 Q & C_{ab}  \\
\hline
b_{[\mu\nu]} & Q^2  &  C_{\mu\nu a b} \\
a_{[\mu\nu]}^+ & \Delta_0+8 +4\sqrt{\Delta_0+4} & C_{\mu\nu} \\
a_{[\mu\nu]}^- & \Delta_0+8 -4\sqrt{\Delta_0+4} &C_{\mu\nu}  
\end{array}
\]
\caption{Masses of the bosonic modes on AdS$_5$ in terms of the
  Laplacian eigenvalues of the internal wavefunctions. 
  $\Delta_0$,
  $\Delta_1$ and $\Delta_L$ are the eigenvalues of the Laplacian on
  scalars, one-forms, and traceless symmetric modes, respectively. $Q$
  is the  eigenvalue of $\imath \star d$. 
The indices $\mu,\nu,\ldots$ are for AdS$_5$, and $a,b,\ldots$ are for the internal manifold.
We set $R_{ab}=4g_{ab}$ for simplicity. 
Other symbols are explained in the main text.\label{tab:masses}}
\end{table}

First, let us recall the well-known relation between the mass eigenvalues of the Kaluza-Klein modes and the Laplacian eigenvalues of the internal wavefunctions on an Einstein 5-manifold, see Table~\ref{tab:masses},  taken from \cite{Ceresole:1999zs,Ceresole:1999ht} and  section 3.6 of \cite{Larsson2004}; we take the standard normalization $R_{ab}=4g_{ab}$.
In the table, the names of the AdS$_5$ modes follow those used in \cite{Kim:1985ez,Ceresole:1999zs,Ceresole:1999ht}.  Here
$\Delta_0$, $\Delta_1$ and $\Delta_L$ are the eigenvalues of the Laplacian on scalars, one-forms, and traceless symmetric modes respectively. $Q$ is the  eigenvalue of $\imath \star d$.\footnote{We use $\imath$ for the unit imaginary number and $\star$ for the Hodge star.}
The 10d fields are the metric $g_{MN}$, the axiodilaton $\imath e^\varphi+ C$, the combined two-form $C_{MN}=B_{MN}^\text{NSNS}+ \imath C_{MN}^\text{RR}$, and the potential $C_{MNRS}$ of the self-dual five-form.
Note that the axiodilaton and the combined two-form are complex fields.
Therefore, for a scalar eigenfunction $f$ on $Y$, the AdS$_5$ scalars $\pi(f)$ and $\pi(f^*)$ are complex conjugates of each other, but $B(f)$ and $B(f^*)=[B^*(f)]^*$ are independent fields, etc. Note also that the on-shell components of a massive two-form field $B_{\mu\nu}$ on 5d spacetime split into two little group multiplets, with spins $(1,0)$ and $(0,1)$. We denote them $B_{\alpha\beta}$ and $B_{\dot\alpha\dot\beta}$.

\subsection{Expansion on Sasaki-Einstein manifolds}\label{KKmodes}
We now restrict our attention to Sasaki-Einstein manifolds. Then, the bosonic modes listed in Table~\ref{tab:masses} should be organized into supermultiplets. To analyze these structures, we first need to recall the geometry of the Sasaki-Einstein manifolds. 

Let $X$ be the Calabi-Yau cone, and take the Sasaki-Einstein manifold $Y$ to be the locus $\rho=1$, where $\rho$ is the radial distance from the tip of the cone.  Using the complex structure $\cI$ on $X,$ we can construct a Killing vector $\xi=\cI \left( \rho\partial_\rho \right)$ on $Y$ called the Reeb vector. 
The rescaled Reeb vector $(2/3)\xi$ is the generator of the R-charge. 
Let $\eta$ be the one-form obtained by contracting $\xi$ with the metric. Define the two-form $J$ via \begin{equation}
d\eta=2J,
\end{equation}
this two-form $J$ is the restriction of the K\"ahler form on the cone $X$ to the base $Y$. Similarly, there is a covariantly-constant $(3,0)$-form $\Omega_{CY}$ on $X$. From this, we can define a $(2,0)$-form $\Omega$ on $Y$, satisfying \begin{equation}
d\Omega= 3\imath\eta \wedge \Omega, \quad
d\bar\Omega=-3\imath\eta\wedge\bar\Omega.
\end{equation}

We can restrict the complex structure $\cI$ of the cone $X$ to the sub-bundle of $TY$ perpendicular to $\eta$; we still denote it by $\cI$. This determines the so-called CR structure on the Sasaki-Einstein manifold. Using this, we can split the complexified tangent bundle locally as \begin{equation}
T_Y = \bC\xi \oplus T^{(1,0)}Y \oplus T^{(0,1)}Y
\end{equation} where $T^{(1,0)}Y$ is the eigenspace on which $\cI$ acts by $\imath$, and $T^{(0,1)}$ is its conjugate. 
The one-forms split accordingly, \begin{equation}
\Omega_Y = \bC\eta \oplus \Omega^{(1,0)}Y \oplus \Omega^{(0,1)}Y
\end{equation}
and therefore we can split the exterior derivative as \begin{equation}
d= \eta \wedge \pounds_\xi + \partial_B +\bar \partial_B. 
\end{equation} The operator $\bar \partial_B$ is called the tangential Cauchy-Riemann operator. It satisfies \begin{equation}
\bar\partial_B^2 = \partial_B^2=0,\qquad
\bar\partial_B\partial_B + \partial_B \bar\partial_B = -2 J\wedge \pounds_\xi.
\end{equation}

For a holomorphic vector bundle $V$ (in the tangential Cauchy-Riemann sense), we can then consider the sections of the bundle \begin{equation}
V\otimes \oplus_q \Omega^{(0,q)}Y
\end{equation} on which $\bar\partial_B$ naturally acts. The cohomology of this complex is called the Kohn-Rossi cohomology of $V$, and is denoted by $H^q_{\tgdb}(Y,V)$. For $V=\Omega^{(p,0)}Y$, this is abbreviated as $H^{p,q}_\tgdb(Y).$  For details, see~e.g.~\cite{MR0177135,MR604043}.

Recall that we split the holomorphic tangent bundle of $X$ as $T_{\bC}X=T'_X\oplus \bC D,$ where $D$ is the holomorphic part of the dilatation vector field on the cone. 
Similarly, we defined $\Omega'_X$ to be the subspace of $\Omega X$ perpendicular to $D$.
 The restrictions of the bundles $T'_X$ and $\Omega'_X$ to the base $Y$ are $T^{(1,0)}Y$ and $\Omega^{(1,0)}Y$ respectively, as defined above. Therefore,  there is a natural map $H^{q}(X,\wedge^p \Omega'_X)\to H^{p,q}(Y)$ given by the restriction. 
Conversely, for an element $\omega\in H^{(p,q)}(Y)$ of R-charge $r$, 
we can define an element $\rho^{3r/2}\omega \in H^q(X,\wedge^p\Omega'_X)$; note that our $X$ does not contain the tip, so we can multiply by $\rho^{3r/2}$ without  problem.
Thus we have an isomorphism between these two linear spaces.

After these preparations, we will now discuss the scalar, vector, antisymmetric and the symmetric traceless modes in turn.
We will find that most of the modes can be constructed from the scalar eigenfunctions. 
We only state the results in this section; the details for the scalar, vector, and two-form modes are given in appendix~\ref{supergravitydetails}. We have not completed the analysis of the symmetric traceless modes. 
Similar analysis for four-dimensional K\"ahler-Einstein spaces was performed by Pope \cite{Pope:1982ad}.

\subsubsection{Scalar eigenfunctions}
The scalar Laplacian $\Delta_0$ and the R-charge operator  $(2/3)\pounds_\xi$ commute with each other, so we can choose simultaneous scalar eigenfunctions $f$.  We abuse the notation and denote the eigenvalue of the scalar Laplacian $\Delta_0$ by $\Delta_0$.  We normalize the $r$-charge of $f$ by $(2/3)\pounds_\xi f = \imath r f$ where $r$ is a real number. Obviously, $f^*$ has the same eigenvalue, $\Delta_0,$ but has the eigenvalue $-r$ under $(2/3)\pounds_\xi$. It turns out to be useful to introduce a positive number $E_0$ satisfying $\Delta_0=E_0(E_0+4)$.
Then there is an eigenvalue bound \begin{equation}
E_0 \ge \frac32r,
\end{equation} which is saturated if and only if $\bar\partial_B f=0$. This happens if and only if $f$ is a restriction of a holomorphic function on the Calabi-Yau cone $X$ to the Sasaki-Einstein base $Y$. 
See Appendix~\ref{sec:details__scalar_eigenfunctions} for details.

\subsubsection{Vector eigenfunctions}
We now analyze the one-form eigenfunctions of the one-form Laplacian.
Given a scalar eigenmode $f$ with the R-charge $r$, consider the one-form modes \begin{equation}
f\eta, \quad  \partial_B f,  \quad \bar\partial_B f, \label{set1}
\end{equation} and the modes \begin{equation}
df . \Omega = \bar\partial_B f . \Omega, \qquad
df . \bar\Omega = \partial_B f . \bar\Omega  \label{set2}
\end{equation}  where the contraction $x.y$ of a one-form $x$ and a two-form $y$ is defined to be $x_\mu g^{\mu\nu} y_{\nu\rho}$. 
Out of the three modes in \eqref{set1}, one linear combination is a gauge mode, the other two are eigenmodes with eigenvalues \begin{equation}
\Delta_1= E_0(E_0+2),\quad  (E_0+2)(E_0+4). 
\end{equation} These two modes have R-charge $r$. 
We  denote the corresponding eigenmodes schematically by $(f\eta)^-$ and $(f\eta)^+$ respectively. When $\bar\partial_B f=0$ we have $(f\eta)^-=\partial_B f$, and $(f\eta)^+=0$. 
The two modes \eqref{set2} are automatically eigenmodes themselves, with eigenvalues \begin{equation}
\Delta_1 = (E_0+1)(E_0+3),
\end{equation} and R-charges $r\pm2$. 

A one-form eigenmode $\nu$ orthogonal to the modes in \eqref{set1} is a co-closed section of $\Omega^{(1,0)}Y \oplus \Omega ^{(0,1)} Y$.  If $v$ is furthermore orthogonal to the modes in \eqref{set2}, either $v$ is a section of $\Omega^{(1,0)}Y$ closed under  $\partial_B$ or a section of $\Omega^{(0,1)}Y$ closed under $\bar\partial_B$, i.e. \begin{equation}
\nu\ \text{or} \ \nu^*\in H^{0,1}_{\tgdb}(Y).
\end{equation}
But this cohomology group is known to be empty, as $H^1(\cO_X)$ vanishes. 
We conclude that  any vector eigenmode is either in \eqref{set1} or in  \eqref{set2}.

Note that a holomorphic vector field $v$ can be thought of as a one-form via $\nu=v\interior \Omega$ satisfying $\bar\partial_B\nu=0$. From the discussions above, there is a scalar function $f$ such that $\nu$ is given either by $\partial_B f$ in \eqref{set1} or $\bar\partial_B f.\Omega$ in \eqref{set2}. The former is impossible. Therefore, any holomorphic vector field $v$ has the form \begin{equation}
v\interior \Omega = \bar\partial_B f.\Omega.\label{holomorphicvec}
\end{equation}
Note that this relation allows us to write down explicit scalar eigenfunctions of the Laplacian of $Y^{p,q}$ and $L^{a,b,c}$, starting from known holomorphic vector fields.
For the $Y^{p,q}$ and $L^{a,b,c}$ manifolds these explicit scalar eigenfunctions were previously identified in \cite{Kihara:2005nt,Oota:2005mr}.
More details of the calculations in this section can be found in Appendix \ref{sec:details__vector_eigenfunctions}.
\subsubsection{Two-form eigenfunctions}
Now we construct two-form eigenfunctions in terms of scalar eigenmodes.
For two-forms, it is convenient to use the operator $\imath\star d$, which satisfies $(\imath\star d)(\imath\star d)=\Delta_2$. 
We denote the eigenvalue of $\imath\star d$ by $Q$. 
Given a scalar eigenmode $f$ with $\Delta_0=E_0(E_0+4)$ and the R-charge $r$ as before, we first consider modes \begin{equation}
df \wedge \eta, \quad f  J, \quad \partial_B \bar \partial_B f \label{Set1}.
\end{equation}  One linear combination is a gauge mode, the other two linear combinations give eigenmodes with \begin{equation}
Q=\pm (E_0+ 2),
\end{equation} and R-charge $r$.  
We denote these eigenmodes schematically by $(fJ)^\pm$. 

Next, consider modes \begin{equation}
f \Omega, \quad \partial_B \bar\partial_B f . \Omega, \quad
\eta(\bar\partial_B f.\Omega), \quad
\bar\partial_B(\bar\partial_B f.\Omega),
\label{Set2}
\end{equation} where $x.y$ for two two-forms stand for
$x_{\mu[\nu}y_{\rho]\sigma}g^{\mu\sigma}$. As shown in the appendix,
$\tgd\tgdb f.\Omega$ is linearly dependent on the others. The
remaining three modes give two eigenmodes with \begin{equation}
Q=E_0+3, \quad -E_0-1, 
\end{equation}  and R-charge $r+2$. We denote these eigenmodes by $(f\Omega)^+$ and $(f\Omega)^-$.
When $\bar\partial_B f.\Omega$ corresponds to a holomorphic vector field as in \eqref{holomorphicvec}, the eigenmode with the eigenvalue $E_0+3$ disappears. When $f$ is holomorphic, both modes disappear.  

Similarly, the modes \begin{equation}
f \bar\Omega, \quad \partial_B \bar\partial_B f . \bar\Omega, \quad
\eta(\partial_B f.\bar\Omega), \quad
\partial_B(\partial_B f.\bar\Omega)
\label{Set3}
\end{equation} give two eigenmodes \begin{equation}
Q=E_0+1, \quad -E_0-3
\end{equation} with the R-charge $r-2$. 
We denote the eigenmodes by $(f\bar\Omega)^+$ and $(f\bar\Omega)^-$.
When $\partial_B f.\bar\Omega$ corresponds to an anti-holomorphic vector field as in \eqref{holomorphicvec}, the eigenmode with the eigenvalue $-E_0-3$ disappear. 
When $f$ is antiholomorphic, both modes disappear.

We see that the modes listed in \eqref{Set1}, \eqref{Set2} and \eqref{Set3} include all two-forms of the form \begin{equation}
\partial_B v,\quad 
\bar\partial_B v,\quad
\partial_B v.\Omega,\quad
\partial_B v.\bar\Omega,\quad
\bar\partial_B v.\Omega,\quad
\bar\partial_B v.\bar\Omega 
\end{equation} for all one-form modes $v$ together with \begin{equation}
fJ, \quad f\Omega, \quad f\bar\Omega
\end{equation} for all scalar modes $f$.  Therefore, a two-form $\omega$ orthogonal to all the modes in \eqref{Set1}, \eqref{Set2} and \eqref{Set3} is a co-closed section of $\Omega^{(1,1)}Y$ which is closed either under $\partial_B$ or $\bar\partial_B$. 
Equivalently, \begin{equation}
\omega\ \text{or}\ \omega^*\in H^{1,1}_{\bar\partial_B}(Y)
\end{equation} If the first of these possibilities is realized, we have \begin{equation}
Q=\frac32 r.
\end{equation} If we further impose $r=0$, $\omega$ is in the ordinary second cohomology $H^2(Y)$. 
Further details of the calculations in this section are given in appendix~\ref{sec:details__2-form_eigenfunctions}.\footnote{These modes were also constructed in Sec.~3.3 and Appendix A of \cite{Baumann:2010sx}.}

\subsubsection{Symmetric traceless eigenfunctions}
Finally, let us consider the symmetric traceless deformation $\delta g_{\mu\nu}$ of the metric tensor.
For each $2$-form $\omega$ with $Q=E$ and R-charge $r$, we find\footnote{Based on the explicit expansions given for $Y=S^5$ in \cite{Kim:1985ez} and for $Y=T^{1,1}$ implicitly given in \cite{Ceresole:1999zs,Ceresole:1999ht} and kindly provided explicitly by Professor Gianguido Dall'Agata to the authors. A general direct analysis is in progress. This structure can also be deduced by demanding that all the bosonic modes fit in the superconformal multiplets correctly, as tabulated in section\ref{tons-of-tables}.  } three symmetric traceless modes we schematically denote by
\begin{equation}
\omega.\Omega,\quad
\omega.J,\quad
\omega.\bar\Omega.
\end{equation} They have eigenvalues \begin{equation}
(E+1)(E-3)+8,\quad
(E+2)(E-2)+8,\quad
(E+3)(E-1)+8
\end{equation} and the R-charges \begin{equation}
r+2,\quad r,\quad  r-2.
\end{equation} 

When $\omega$ is itself constructed from a scalar $f$ as in the previous subsection, 
there are some overlaps in this construction such as $(f\Omega)^+.\bar\Omega \propto (f\bar\Omega)^-.\Omega$ and $(f\Omega)^+.\Omega \propto (f\Omega)^-.\Omega$.
In the end, we find nine metric modes constructed from $f$, which are \begin{equation}
\begin{array}{ccc}
(f\Omega)^+.\Omega, & (fJ)^+.\Omega, & (f. \bar\Omega)^+.\Omega, \\
(f\Omega)^+. J, & (fJ)^+.J, & (f\bar\Omega)^+.\Omega, \\
(f\Omega)^+. \bar\Omega, & (fJ)^+.\bar\Omega, & (f.\bar\Omega)^+. \bar\Omega.
\end{array} 
\end{equation}
whose eigenvalues under the Lichnerowicz Laplacian are given by \begin{equation}
\begin{array}{rrr}
E_0(E_0+4)+8, & (E_0-1)(E_0+3)+8, & (E_0-2)(E_0+2)+8,\\
(E_0+1)(E_0+5)+8, & E_0(E_0+4)+8, &  (E_0-1)(E_0+3)+8,\\
(E_0+2)(E_0+6)+8, & (E_0+1)(E_0+5)+8, & E_0(E_0+4)+8.
\end{array}
\end{equation} and the R-charges are \begin{equation}
\begin{array}{lll}
r+4, & r+2, & r, \\
r+2, & r, & r-2, \\
r, & r-2, &r-4.
\end{array}
\end{equation}

We have not directly checked that there are no other eigenmodes. However, the fact that the modes found so far can be fit into supermultiplets implies that there cannot be any other modes.

\subsection{Supermultiplet structures on Sasaki-Einstein manifolds}\label{tons-of-tables}
The mass eigenvalues of various Kaluza-Klein modes are given by feeding the Laplacian eigenvalues obtained in section \ref{KKmodes}
to the relations given in Table~\ref{tab:masses}.  The Kaluza-Klein modes nicely arrange into superconformal multiplets which we now describe.
The multiplets containing modes constructed from a scalar eigenmode $f$ with $\Delta_0=E_0(E_0+4)$ with the R-charge $r$ is listed in Tables~\ref{graviton},
\ref{gravitino1}, \ref{gravitino2},
\ref{vector1}, \ref{vector2}, \ref{vector3}.
We call them the ``graviton multiplets'', the ``gravitino multiplets I, II, III, IV'', and the ``vector multiplets I, II, III, IV'' following \cite{Ceresole:1999zs,Ceresole:1999ht}. The name refers to the top component of the supermultiplet \emph{when it is not shortened}. 
For the particular case of $T^{1,1}$, our tables reproduce theirs.\footnote{Note that the tables in \cite{Ceresole:1999zs,Ceresole:1999ht} contain typos, as already pointed out in Sec.~5.2 of \cite{Baumann:2010sx}. The authors thank Professor Gianguido Dall'Agata for correspondences concerning this point.}

\begin{table}
\[
\begin{array}{ccc|c|c|l|l}
&&&\text{mode} & \text{w.f.} & \text{dim} & \text{R}  \\
\hline
\diamond&\star&\bar\star&H_{\mu\nu} & f & E_0+4 & r \\
\diamond&\star&\bar\star&\phi_\mu & (f\eta)^- & E_0+3 & r \\
&\star&&a_\mu & (\bar\partial_B f).\Omega & E_0+4 & r+2 \\
&&\bar\star&a_\mu^* & (\partial_B f).\bar\Omega & E_0+4 & r-2 \\
&&&B_\mu & (f\eta)^+ & E_0+5 & r \\
&\star&&b_{\alpha\beta}^+ & (fJ)^+ & E_0+4 & r \\
&&\bar\star&b_{\dot\alpha\dot\beta}^- & (fJ)^- & E_0+4 & r \\
&&&\phi &  (fJ)^+.J & E_0+4 & r
\end{array}
\]
\caption{The  ``graviton multiplet''. 
When conserved, the lowest component is $\phi_\mu$.
The symbols $\diamond$, $\star$ and $\bar\star$ denote the components which remain when $f$ is a constant, holomorphic, and antiholomorphic, respectively. \label{graviton}}
\end{table}

\begin{table}
\[
\begin{array}{ccc|c|c|l|l}
&&&\text{mode} & \text{w.f.} & \text{dim} & \text{R}  \\
\hline
\diamond&\bullet&\star& \phi_\mu & (\bar\partial_B f).\Omega &  E_0+2 & r+2 \\
&&\star& a_\mu & (f\eta)^- & E_0+3 & r \\
\diamond&\bullet &\star & a_{\alpha\beta}^- & f & E_0+2 & r \\
&\bullet& & b_{\alpha\beta}^- & (f\Omega)^- & E_0+3 & r+2 \\
& &\star & a & (fJ)^- &   E_0+2 & r \\
&&& \phi & (fJ)^+.\Omega & E_0+3  & r +2 \\
\hline
\bar\diamond&\bar\bullet&\bar\star& \phi_\mu & (\partial_B f).\bar\Omega &  E_0+2 & r-2 \\
&&\bar\star& a_\mu^* & (f\eta)^- & E_0+3 & r \\
\bar\diamond&\bar\bullet &\bar\star & a_{\dot\alpha\dot\beta}^- & f & E_0+2 & r \\
&\bar\bullet& & b_{\dot\alpha\dot\beta} & (f\bar\Omega)^- & E_0+3 & r-2 \\
& &\bar\star & a & (fJ)^+ &   E_0+2 & r \\
&&& \phi & (fJ)^+.\bar\Omega & E_0+3  & r -2
\end{array}
\]
\caption{The  ``gravitino multiplet I'' (top) and its CP conjugate the  ``gravitino multiplet III''. 
When long, the lowest component has spin $(1/2,0)$, with dimension $E_0+3/2$ and the R-charge $r+1$.
The symbols $\bullet$ and $\star$ mark the components which remain when $f$ is a holomorphic function and when $(\bar\partial_B f).\Omega$ is a holomorphic vector, respectively. The symbol $\diamond$ is when $f$ is holomorphic and of dimension 1. Then $\phi_\mu$ becomes massless with non-zero R-charge, signifying the enhancement of the supersymmetry.  \label{gravitino1}}
\end{table}

\begin{table}
\[
\begin{array}{c|c|c|l|l}
&\text{mode} & \text{w.f.} & \text{dim} & \text{R}  \\
\hline
\star & a_\mu & (f\eta)^+ & E_0+5 & r \\
& B_\mu & (\bar\partial_B f).\Omega &  E_0+6 & r+2 \\
\star & b_{\alpha\beta} & (f\Omega)^+ & E_0+5 & r+2 \\
\star & a_{\alpha\beta}^+ & f & E_0+6 & r \\
& \phi & (f\Omega)^+.J & E_0+5  & r +2\\
&  a & (fJ)^+ &   E_0+6 & r \\
\hline
\bar\star & a_\mu & (f\eta)^+ & E_0+5 & r \\
& B_\mu & (\partial_B f).\bar\Omega &  E_0+6 & r-2 \\
\bar\star & b_{\dot\alpha\dot\beta} & (f\bar\Omega)^+ & E_0+5 & r-2 \\
\bar\star & a_{\dot\alpha\dot\beta}^+ & f & E_0+6 & r \\
& \phi & (f\bar\Omega)^+.J & E_0+5  & r -2\\
&  a & (fJ)^+ &   E_0+6 & r \\
\end{array}
\]
\caption{The  ``gravitino multiplet II'' (top) and its CP conjugate the`` gravitino multiplet IV''.
The lowest component has spin $(1/2,0)$, with dimension $E_0+9/2$ and the R-charge $r+1$.  The symbols $\star$, $\bar\star$ mark the components which remain when $f$ is holomorphic or anti-holomorphic, respectively.
\label{gravitino2}}
\end{table}

\begin{table}
\[
\begin{array}{ccccc|c|c|l|l}
&&&&&\text{mode} & \text{w.f.} & \text{dim} & \text{R}  \\
\hline
\diamond &&\istar&&\bar\istar& \phi_\mu & (f\eta)^- & E_0+1 & r \\
\diamond &\star&\istar&\bar\star&\bar\istar& b & f & E_0 & r \\
&\star&\istar &&& a & (f\bar\Omega)^- & E_0+1 & r-2 \\
&&&\bar\star&\bar\istar& a & (f\Omega)^- & E_0+1 & r+2 \\
&&&&& \phi &  (f\bar\Omega^+).\Omega & E_0+2 & r 
\end{array}
\]
\caption{The  ``vector multiplet I''.
The symbols $\diamond$, $\star$, $\istar$, $\bar\star$, $\bar\istar$ denote the components which survive when $f$ generates a Killing vector, when $f$ is holomorphic, when $\bar\partial_B f.\Omega$ is a holomorphic vector, 
when $f$ is anti-holomorphic or when $\partial_B f.\Omega$ is an antiholomorphic vector, respectively.
\label{vector1}}
\end{table}

\begin{table}
\[
\begin{array}{c|c|l|l}
\text{mode} & \text{w.f.} & \text{dim} & \text{R}  \\
\hline
B_\mu & (f\eta)^+ & E_0+7 & r \\
\phi &  (f.\Omega)^+.\bar\Omega & E_0+6 & r \\
a & (f\Omega)^+ & E_0+7 & r+2 \\
a & (f\bar\Omega)^+ & E_0+7 & r-2 \\
\pi & f & E_0+8 & r 
\end{array}
\]
\caption{The  ``vector multiplet II''.  This multiplet never shortens.  \label{vector2}}
\end{table}

\begin{table}
\[
\begin{array}{cc|c|c|l|l}
&&\text{mode} & \text{w.f.} & \text{dim} & \text{R}  \\
\hline
&\bar\star&a_\mu & (\partial_B f).\bar\Omega & E_0+4 & r-2 \\
\bar\bullet&\bar\star&a &  (f\bar\Omega)^-  & E_0+3 & r-2 \\
\bar\bullet&\bar\star&B & f & E_0+4 & r \\
&&\phi & (f\bar\Omega)^+.\bar\Omega & E_0+4 & r-4 \\
&&a &  (f\bar\Omega)^+ & E_0+5 & r-2 \\
\hline
&\star&a_\mu & (\bar\partial_B f).\Omega & E_0+4 & r+2 \\
\bullet&\star&a &  (f\Omega)^-  & E_0+3 & r+2 \\ 
\bullet&\star&B & f & E_0+4 & r \\
&&\phi & (f\Omega)^+.\Omega & E_0+4 & r+4 \\
&&a^* &  (f\Omega)^+ & E_0+5 & r+2 
\end{array}
\]
\caption{The  ``vector multiplet III'' (top) and its CP conjugate the ``vector multiplet IV''.
The symbols $\bullet$, $\bar\bullet$, $\star$ and $\bar\star$ mark the components that survive when $f$ is holomorphic, when $f$ is anti-holomorphic, when $\bar\partial_B f.\Omega$ is a holomorphic vector, or when $\partial_B f.\bar\Omega$ is an antiholomorphic vector, respectively.
\label{vector3}}
\end{table}

\begin{table}
\[
\begin{array}{c|c|l|l}
\text{mode} & \text{w.f.} & \text{dim} & \text{R}  \\
\hline
a & \omega &   E_0 & r \\
\phi & \omega.\bar\Omega & E_0+1  & r-2 \\
\hline
b_{\alpha\beta}^- & \omega & E_0+2 & r \\
\phi & \omega &   E_0+2 & r \\
\hline
\phi & \omega.\Omega & E_0+3  & r+2\\
a^* & \omega &   E_0+4 & r \\
\end{array}
\]

\caption{The  special multiplets constructed from $\omega$. The second one has the lowest component with spin $(1/2,0)$, dimension $E_0-1/2$, R-charge $r+1$.  \label{specialNinjas}}
\end{table}

\begin{table}
\[
\begin{array}{c|c|l|l}
\text{mode} & \text{w.f.} & \text{dim} & \text{R}  \\
\hline
B_{\mu} & \omega & 3  & 0\\
\phi & \omega &   2 & 0 \\
\hline
\phi & \omega.\Omega & 3  & 2\\
a & \omega &   4 & 0 \\
\hline
\phi & \omega.\bar\Omega & 3  & -2\\
a^* & \omega &   4 & 0 \\
\end{array}
\]
\caption{The Betti multiplets; note that $\omega=\omega^*$, $r=0$, $E_0=0$.   \label{Betti}}
\end{table}

The modes constructed from $\omega \in H^{1,1}_{\tgdb}(Y)$ with $E_0=Q=(3/2)r$  
 are in the multiplets shown in Table~\ref{specialNinjas}.  
 Note that the KK modes in these three multiplets are complements of the shortened multiplets in ``vector multiplet I'' of Table~\ref{vector1},  in ``gravitino multiplet I'' of Table~\ref{gravitino1} and ``vector multiplet IV'' of Table~\ref{vector2}, and  respectively. There are of course three CP conjugate multiplets constructed from $\bar\omega$.  
 
 For $\omega\in H^2(Y)$, we   have the ``Betti'' multiplets given in Table~\ref{Betti}. Formally, they are  obtained by setting $r=0$ to the modes in Table~\ref{specialNinjas}, but the outcome is quite different. 
The mode $B_\mu$ purely comes from the 4-form: $C_{\mu abc}=B_\mu\wedge (\star \omega)_{abc}$. This is the same mode as  $b^{-}_{\mu\nu}$ with the internal wavefunction $\omega$, appearing in Table~\ref{specialNinjas}; recall that a massless two-form potential is  electro-magnetically dual to a massless one-form potential in five dimensions. Also, the choice of the two branches of $E$ given by solving $\Delta_0=E(E+4)$ needs to be different from those in Table~\ref{specialNinjas}.

\subsection{Short multiplets contributing to the index}
After all these labors, we can now enumerate short multiplets contributing to the superconformal index and compare them with the gauge theory. First, for each non-constant holomorphic function $f$, we find (cf. section \ref{op-hol})
\begin{itemize}
\item A chiral scalar in the ``vector multiplet I'' marked by $\bullet$ in Table~\ref{vector1},
 identifiable with $\tr \cO_f$,
\item A chiral spinor in the ``gravitino multiplet I'' marked by $\bullet$ in Table~\ref{gravitino1}, 
identifiable with $\tr W_\alpha \cO_f$,
\item Another chiral scalar in the ``vector multiplet IV'' marked by $\bullet$ in Table~\ref{vector3},
 identifiable with $\tr W_\alpha W^\alpha \cO_f$,
\item A semiconserved spinor in the ``gravitino multiplet III''  marked by $\star$  in Table~\ref{gravitino1}, 
identifiable with $\tr \bar W_{\dot\alpha} \cO_f$,
\item A semiconserved vector in the ``graviton multiplet''  marked by $\star$  in Table~\ref{graviton}, 
identifiable with $\tr \bar W_{\dot\alpha} W_\alpha \cO_f$,
\item Another semiconserved spinor in the ``gravitino multiplet IV'' marked by $\star$ in Table~\ref{gravitino2}, 
identifiable with $\tr \bar W_{\dot\alpha} W_\alpha  W^\alpha \cO_f$. 
\end{itemize}
They are the modes associated to nonconstant elements in $H^0(X,\cO_X)$ and $H^0(X,\wedge^2 \Omega'_X)$.
Together, they contribute $t^{3r}+t^{3r+6}$ to the index. 

When $f=1$, we only include the third, the fifth, and the sixth modes from the sextuple above, because the first, the second and the fourth are singletons and correspond to a decoupled U$(1)$ multiplet. 
Together, the contribution to the index is $t^6$.

Second, for each holomorphic vector $v$ with R-charge $r$ that comes from a non-holomorphic scalar $f$ as in \eqref{holomorphicvec}, we find (cf. section \ref{op-vec})
\begin{itemize}
\item A semiconserved scalar in the ``vector multiplet I'' marked by $*$ in Table~\ref{vector1},
identifiable with $\tr \cO_v$,
\item A semiconserved spinor in the ``gravitino multiplet I'' marked by $\star$ in Table~\ref{gravitino1},
identifiable with $\tr W_\alpha \cO_v$,
\item Another semiconserved scalar in the ``vector multiplet IV'' marked by $\star$ in Table~\ref{vector3},
identifiable with $\tr W_\alpha W^\alpha \cO_v$.
\end{itemize}
These are the modes associated to $H^0(X,\Omega'_X)$. Together, they contribute $-t^{3r}$ to the index. 

We also have the modes listed in Table~\ref{specialNinjas} which come from two-forms in $\omega=H^{1,1}(Y)\simeq H^1(X,\Omega'_X)$.
For each $\omega$, we find three short multiplets whose charges match those of  $\tr \cO_w$, $\tr W_\alpha \cO_w$ and $\tr W_\alpha W^\alpha \cO_w$ for some word $\cO_w$. Then there is again a cancellation of the contributions from insertions of $W_\alpha$  and of spacetime derivatives to the superconformal index. 

Finally we have the modes listed in Table~\ref{Betti} which come from two-forms in the ordinary second cohomology, $H^2(Y)$. 
For each such two-form $w$, we have a conserved current for a baryonic symmetry and a exactly marginal chiral scalar. The contributions to the index from these modes cancel out. Thus we confirm that the supergravity analysis and the gauge theory analysis fully agree.

\section{Conclusions}\label{conclusions}

We have examined the single-trace superconformal index of the gauge theory on the D3-branes probing a Calabi-Yau cone $X$ using both gauge theory and supergravity. 
On the gauge theory side, we have a quiver gauge theory, whose index can be calculated from the determinant of a matrix $\chi(t)$ encoding the quiver diagram. Utilizing the gauge theory's relation to Ginzburg's dg algebra, we showed that the superconformal index is given by 
\begin{equation}
\sum_{0\le p-q\le 2} (-1)^{p-q} \Tr  t^{3R} \bigm| H^q(X,\wedge^p \Omega'_X) .
\end{equation}

On the supergravity side, we performed the Kaluza-Klein expansion of type IIB supergravity fields on the Sasaki-Einstein base $Y$, and found that the index is given by 
 \begin{equation}
\sum_{0\le p-q\le 2} (-1)^{p-q} \Tr  t^{3R} \bigm| H^{p,q}_{\tgdb}(Y) .
\end{equation}  The equality of the two expressions follows from the fact that an element of $H^{p,q}_{\tgdb}(Y)$ is given by a restriction of an element of $H^q(X,\wedge^p \Omega'_X)$ to $Y$.

In our paper we assumed the Sasaki-Einstein manifold $Y$ to be smooth. It would be interesting to allow orbifold singularities in $Y$ itself, and to see how the analysis is modified. Furthermore, we only considered mesonic operators; it would be interesting to consider (di)baryonic operators involving the determinants in the gauge groups. 

The superconformal index of quivers for toric Calabi-Yau cones were also studied in \cite{Yamazaki:2012cp,Terashima:2012cx} from a rather different perspective. It will be of interest to see what connections, if any, there are with our work. 

One problem we have not been able to answer in general is why the determinant $\det\chi(t)$ factorizes in general: \begin{equation}
\det(\chi(t))= \prod_{i=1}^{n_v} (1-t^{3r_i}).\label{lll}
\end{equation} 
Also, we would like to understand the role the elements  $v_i \in \oplus_{p,q} H^q(\wedge^p \Omega'_X)$ corresponding to the factors in \eqref{lll} play in the physics and the mathematics of the quiver gauge theory.
We observe that the vector space \begin{equation}
 \oplus_{p,q} H^q(\wedge^p \Omega_X)   \simeq \oplus_{p,q} H^q(\wedge^p T_X)  
\end{equation} is the space of states of the closed topological string on $X.$  This space of states was extracted from Ginzburg's dg algebra $\fD$ 
which describes the algebra of open-string states of various D-branes on  $X$, as was proposed in \cite{Kapustin:2004df}.
The factorization seems to arise from the interaction between the open and closed topological strings on $X$.

Another question which deserves to be better understood is the structure of the eigenmodes of $p$-forms and of traceless symmetric tensor fields on the Sasaki-Einstein 5-manifolds. In this paper, they are studied by a laborious, brute-force manipulation, and we found that eigenvalues of various modes are related in a regular, intricate manner. Morally speaking, these relations arise from the fact that each eigenmode can be used as an internal wavefunction for more than one supergravity field, thus giving rise to component fields in more than one  supermultiplet of five-dimensional supergravity.  The actions of the supersymmetry generators on those different multiplets correspond to different geometric operations we can perform on the same eigenmode to produce multiple eigenmodes with Laplacian eigenvalues related to the original one. 
One should be able to distill this structure and express it purely in terms of the geometry of the Sasaki-Einstein manifold, thus streamlining the analysis of this paper.  Of course, it would also be nice to study the Kaluza-Klein expansion of the fermionic modes explicitly, and to check that they fit into the supermultiplets we found in this paper. 

We can also endeavor to compare the supergravity and the gauge theory indices for other holographic pairs, such as the large class of 4d $\cN=2$ and $\cN=1$ models based on M-theory compactification on AdS$_5$ \cite{Gaiotto:2009gz,Benini:2009mz,Bah:2012dg}, or even 3d supersymmetric Chern-Simons-matter theories holographically dual to M-theory on AdS$_4$ times Sasaki-Einstein 7-manifolds. 

The authors hope to come back to at least some of these topics in the future. 

\section*{Acknowledgements}
The authors are thankful for helpful discussions with Alexey Bondal, Sergey Galkin, Simeon Hellerman and especially Yu Nakayama.
The authors also thank Gianguido Dall'Agata for helpful correspondences. 
The authors thank Kentaro Nagao and Yukinobu Toda for assistance on mathematical matters.
YT thanks Dario Martelli for informing him originally about the Kohn-Rossi cohomology a few years ago, when they were working on a project which unfortunately remains unfinished.

 This work is  supported in part by World Premier International Research Center Initiative
(WPI Initiative),  MEXT, Japan through the Institute for the Physics and Mathematics
of the Universe, the University of Tokyo. JS's work is supported by the Japan
Society for the Promotion of Science (JSPS).

\begin{appendix}


\section{Details of the supergravity calculation}\label{supergravitydetails}

\subsection{Conventions}
\label{sec:supergravitydetails_conventions}

We give detailed derivations complementing the material presented in section
\ref{supergravity}. Before doing so, we list our conventions. The
Hodge star, the adjoint to the exterior derivative, and Hodge Laplacian are defined via
\begin{equation}\label{eq:geometry_conventions}
  \begin{aligned}
    \star \omega_{\mu_1 \dots \mu_{d-p}} &= \frac{\sqrt{g}}{p!}
    \epsilon_{\mu_1 \dots \mu_{d-p}}^{\phantom{\mu_1 \dots
        \mu_{d-p}}\nu_1 \dots \nu_p} \omega_{\nu_1 \dots \nu_p}, \\
    \delta \omega_{\mu_1 \dots \mu_{p-1}} &= - \nabla^{\mu_0}
    \omega_{\mu_0 \dots \mu_{p-1}}, \\
    \Delta &= \delta d + d \delta.
  \end{aligned}
\end{equation}
Imposing transverse gauge, $\delta \omega = 0$, the Hodge Laplacian
takes the form
\begin{equation}
  \Delta \omega = - (p+1) \nabla^{\mu_0} \nabla_{\lbrack \mu_0}
  \omega_{\mu_1 \dots \mu_p \rbrack} dx^1 \otimes \dots \otimes dx^{p}.
\end{equation}
The curvature tensor satisfies
\begin{equation}
  \begin{aligned}
    R^\kappa_{\phantom{\kappa}\lambda\mu\nu} &= \partial_\mu
    \Gamma^\kappa_{\nu\lambda} - \partial_\nu
    \Gamma^\kappa_{\mu\lambda} + \Gamma^\kappa_{\mu\rho}
    \Gamma^\rho_{\nu\lambda} - \Gamma^\kappa_{\nu\rho} \Gamma^\rho_{\mu\lambda},
  \end{aligned}
\end{equation}
with the Ricci tensor given by $R_{\mu\nu} =
R^\kappa_{\phantom{\kappa}\mu\kappa\nu}$. 

Let us now turn to some aspects of Sasaki-Einstein
geometry. Quantities on the CY cone are denoted with an $X$,
quantities on the KE base with $KE$, quantities on the
five-dimensional $SE$ come without any modifiers. Due to the Einstein
condition, the Ricci tensor is related to the metric via $R_{\mu\nu} = 4 g_{\mu\nu}$.
Next, note that the
symplectic forms satisfy
\begin{equation}
  \begin{aligned}
    J &= \frac{1}{2} d\eta, \quad
    J_X &= \frac{1}{2} d(\rho^2 \eta).
  \end{aligned}
\end{equation} where $\rho$ is the radial coordinate.
Since $J_X$ is covariantly constant ($\nabla^X J_X = 0$), one finds\footnote{
This makes use of the relation between six- and five-dimensional connection:
\begin{equation*}
  (\Gamma^X)^\rho_{\mu\nu} = - \rho g_{\mu\nu}, \quad
  (\Gamma^X)^\kappa_{\mu \rho} = \rho^{-1} \delta^\kappa_\mu, \quad
  (\Gamma^X)^\kappa_{\mu\nu} = \Gamma^\kappa_{\mu\nu}.
\end{equation*}
}
\begin{equation}
  \begin{aligned}
    \nabla_\kappa J_{\mu\nu} &= - g_{\kappa\mu} \eta_\nu +
    g_{\kappa\nu} \eta_\mu.
  \end{aligned}
\end{equation}

We deal similarly with the holomorphic $(3,0)$ form $\Omega^X$ (again
$\nabla^X \Omega^X = 0$). Decomposing
\begin{equation}
  \frac{\Omega^X}{\rho^3} = \left( \frac{d\rho}{\rho} + \imath \eta \right)
  \wedge \Omega,
\end{equation}
and subsequently expanding $\eta^\lambda (\nabla^X_\kappa
\Omega^X_{\lambda\mu\nu})$ leads to
\begin{equation}\label{eq:SE-identities_20-form_covariant_derivative}
  \nabla_\kappa \Omega_{\mu\nu} = \imath \eta \wedge \Omega_{\kappa\mu\nu}.
\end{equation}
This implies
\begin{equation}
  d\Omega = 3 \imath \eta \wedge \Omega, \quad
  \delta\Omega = 0, \quad
  \pounds_\xi \Omega = 3 \imath \Omega.
\end{equation}

The restriction of the symplectic form to the base satisfies
\begin{equation}
  J^{KE}_{a\bar{b}} = \imath g^{KE}_{a\bar{b}}.
\end{equation}

On K\"ahler manifolds, one can choose the symplectic two-form to be
either self dual or anti-self dual. In the Sasaki-Einstein case, self
duality generalizes to
\begin{equation}\label{eq:Hodge-dual_J}
  \star J = J \wedge \eta. 
\end{equation}
Calculating $\star (J \wedge J)$ one finds that
\begin{equation}\label{eq:Hodge-dual_Reeb}
  \star \eta = \frac{1}{2} J \wedge J.
\end{equation}
Similarly, the $(2,0)$ form satisfies
\begin{equation}\label{eq:Hodge-dual_Omega}
  \star\Omega = \Omega \wedge \eta, \qquad
  \star\eta = \frac{1}{4} \Omega \wedge \bar{\Omega}.
\end{equation}
Since $\star \eta \wedge \eta = \vol_{SE}$, the volume form can be
expressed using $\star 1 = \frac{1}{2} J \wedge J \wedge \eta$.

The tangential Cauchy-Riemann
operators are
\begin{equation}
  \begin{aligned}
    \tgd &= \partial - A^+ \wedge \pounds_\xi,
    \qquad
    \tgdb = \bar{\partial} - A^- \wedge \pounds_\xi.
  \end{aligned}
\end{equation}
They satisfy
\begin{equation}\label{eq:CR_operator--properties}
  \begin{aligned}
    &\tgd \tgdb + \tgdb \tgd = -2 J \wedge \pounds_\xi, &
    \quad
    \tgd \tgd &= 0, \\
    &d = \tgd + \tgdb + \eta \wedge \pounds_\xi, &
    \quad
    \tgdb\tgdb &= 0.
  \end{aligned}
\end{equation}

It is convenient to introduce a series of projection operators,
\begin{equation}\label{eq:SE-conventions__projection_operators}
  \begin{aligned}
    \Pi_\mu^{\text{KE}\nu} &= g_\mu^{\phantom{\mu}\nu} - \eta_\mu
    \eta^\nu = - J_\mu^{\phantom{\mu}\lambda} J_\lambda^{\phantom{\lambda}\nu} , \\
    \Pi^{\pm\nu}_\mu &= \frac{1}{2} (g_\mu^{\phantom{\mu}\nu} \mp \imath
    J_\mu^{\phantom{\mu}\nu} - \eta_\mu \eta^\nu ) = \frac{1}{2}
    (g_\mu^{\phantom{\mu}\lambda} \mp \imath J_\mu^{\phantom{\mu}\lambda}) \Pi_\lambda^{\text{KE}\nu},
  \end{aligned}
\end{equation}
that project onto the K\"ahler-Einstein base and (anti-) holomorphic
indices respectively. 
For index calculations, it can be useful to express forms of definite degree as
\begin{equation}\label{eq:def_degree_forms_and_projections}
  \alpha^{(p,q)} = \frac{1}{p! q!} \Pi_{\mu_1}^{+\nu_1} \dots
  \Pi_{\mu_{p+q}}^{-\nu_{p+q}} \alpha_{\nu_1 \dots \nu_{p+q}} dx^{\mu_1}
  \wedge \dots \wedge dx^{\mu_{p+q}}.
\end{equation}
To give an example, consider the action of $\tgd$ on a $(1,0)$ form $\alpha$:
\begin{equation}
  \tgd\alpha = \frac{1}{2} \Pi_\mu^{+\kappa} \Pi_\nu^{+\lambda}
  d\alpha_{\kappa\lambda} dx^\mu \wedge dx^\nu.
\end{equation}
Also, the projection operators allow us to easily generalize
identities that are more obvious for K\"ahler manifolds:
\begin{equation}\label{eq:omega__omega-bar__identity}
  \bar{\Omega}^{\mu\lambda} \Omega_{\lambda\nu} = -4 \Pi^{-\mu}_{\phantom{-\mu}\nu}.
\end{equation}

\subsection{Details of the calculations}
\label{sec:details-calculations}

To avoid clutter, we frequently denote the eigenvalue of the R-charge
by $q = \frac{3}{2} r$.

\subsubsection{Scalar eigenfunctions}
\label{sec:details__scalar_eigenfunctions}

We start with the bound on scalars. Normalizing $f$ suitably, one finds
\begin{equation}
  \begin{aligned}
    \int \vol \bar{f} \Delta_0 f &= \int \vol \nabla^\mu \bar{f}
    \nabla_\mu f \\
    &= \int \vol \left( 2 g^{a \bar{b}} \tgn_a \bar{f}
      \tgnb_{\bar{b}} f + g^{a\bar{b}} \bar{f} \lbrack \tgn_a, \tgnb_{\bar{b}}
    \rbrack f + \pounds_\xi \bar{f} \pounds_\xi f \right).
  \end{aligned}
\end{equation}
Acting on scalar functions, the commutator evaluates to
$\lbrack \tgn_a, \tgnb_{\bar{b}} \rbrack f = -2 J_{a\bar{b}} \pounds_\xi f$ and since
$\pounds_\xi f = \imath q f$, it follows that
\begin{equation}
  \begin{aligned}
    \int \vol \bar{f} \Delta_0 f &= \int \left\lbrack \vol 2 \vert \tgdb f \vert^2 +
    4 q \bar{f} f + q^2 \bar{f} f \right\rbrack.
  \end{aligned}
\end{equation}
Therefore
\begin{equation}
  \begin{aligned}
    E_0 (E_0 + 4) = \Delta_0 \geq q (q + 4),
  \end{aligned}
\end{equation}
confirming that $E_0 \geq \frac{3}{2} r$ with equality if and only if
$f$ is holomorphic with respect to the CR operator.

\subsubsection{Vector eigenfunctions}
\label{sec:details__vector_eigenfunctions}

While the one-form eigenmodes with R-charge $r$ can be expressed in the
basis \eqref{set1}, the choice
\begin{equation}
  \begin{aligned}
    &v_1 \equiv f \eta, \quad
    v_2 \equiv \imath (\tgd f + \tgdb f) = \imath dx^\mu
    \Pi_\mu^{KE\nu} \partial_\nu f, \\
    &v_3 \equiv \imath (\tgd f - \tgdb f) = dx^\mu
    J_\mu^{\phantom{\mu}\nu} \partial_\nu f,
  \end{aligned}
\end{equation}
is more suitable for calculations. Of course, this basis includes the gauge
mode $df = \imath q v_1 -\imath v_2$. However, not all of $v_i$ are eigenmodes
of the Laplacian. Instead,
\begin{equation}
  \Delta_1 v_i = M_{ij} v_j,
\end{equation}
where
\begin{equation}
  \begin{aligned}
    M_{ij} &= 
    \begin{pmatrix}
      \Delta_0 + 8 & 0 & 2 \\
      8q & \Delta_0 & 2q \\
      2(\Delta_0 - q^2) & 2q & \Delta_0
    \end{pmatrix}.
  \end{aligned}
\end{equation}
The eigenvalues of the Laplacian are thus those of $M$,
\begin{equation}
  \begin{array}{|c|l|l|}
    \hline
        \text{Mode} & \text{Eigenvalue} & \\
    \hline
    df & \Delta_0 & E_0(E_0+4) \\
    f\eta^- & \Delta_0+4 - 2 \sqrt{\Delta_0+4} & E_0(E_0+2) \\
    f\eta^+ & \Delta_0+4 + 2 \sqrt{\Delta_0+4} & (E_0+2)(E_0+4) \\
    \hline
  \end{array}
\end{equation}
with
\begin{equation}
  \begin{aligned}
    f\eta^- &= \imath \frac{(E_0+4)^2}{8q (E_0+2)} \lbrack (E_0^2 -
    q^2) v_1 + q v_2 - E_0 v_3 \rbrack, \\
    f\eta^+ &= -\imath \frac{E_0^2}{8 q (E_0+2)} \{ \lbrack (E_0 +
    4)^2 - q^2 \rbrack v_1 + q v_2 + (E_0+4) v_3 \}.
  \end{aligned}
\end{equation}

Now, holomorphy of $f$ translates to $v_2 = v_3$. $\Delta_1 v_2 =
\Delta_1 v_3$ demands  $E_0 = q$, saturating the bound
derived in the previous section
(\ref{sec:details__scalar_eigenfunctions}). One sees that $f\eta^- =
0$, while
\begin{equation}
  f\eta^+ = -\frac{\imath q}{4} (4 v_1 + v_2) =
  df + \frac{q-4}{4} \tgd f.
\end{equation}

The one-forms
\begin{equation}
  df.\Omega, \qquad
  df.\bar{\Omega}
\end{equation}
are considerably simpler. Using
\eqref{eq:SE-identities_20-form_covariant_derivative} one finds by
direct calculation
\begin{equation}
  \Delta_1 \nabla^\lambda f \Omega_{\lambda\mu} = (\Delta_0 + 3)
  \nabla^\lambda f \Omega_{\lambda\mu}
\end{equation}
and similar for $\bar{\Omega}$. The change in R-charge is similarly
straightforward.

\subsubsection{Two-form eigenfunctions}
\label{sec:details__2-form_eigenfunctions}
In this section, the dot operator is defined as
\begin{equation}
  \begin{aligned}
    \alpha_{(2)}.\beta_{(2)} = \frac{1}{2} \alpha_{\mu\rho}
    \beta^\rho_{\phantom{\rho}\nu} dx^\mu \wedge dx^\nu, \qquad
    \alpha_{(1)}.\beta_{(2)} = \alpha^\lambda \beta_{\lambda\mu}
    dx^\mu.
  \end{aligned}
\end{equation}

\subsubsection*{R-charge $r$}

For the two-forms with R-charge $r$, we choose the following basis:
\begin{equation}
  \begin{aligned}
    &v_1 \equiv \tgd\tgdb f, \quad
    v_2 \equiv df \wedge \eta = (\tgd f + \tgdb f) \wedge \eta, \\
    &v_3 \equiv \imath (df.J)\wedge \eta = (\tgd f - \tgdb f) \wedge\eta.
  \end{aligned}
\end{equation}
Note that $df \wedge \eta = -2 f J + d(f\eta)$. We want to calculate
the eigenvalue of $Q = \imath\star d$.

Making use of \eqref{eq:CR_operator--properties} one finds
\begin{equation}
  \begin{aligned}
    dv_1 &= \imath q ( \tgd\tgdb f \wedge \eta - 2 \tgdb f \wedge J ),
    \\
    dv_2 &= -2 df \wedge J, \\
    dv_3 &= 2 \left\lbrack - \tgd \tgdb f \wedge \eta - \imath q f J \wedge
    \eta - (\tgd f - \tgdb f) \wedge J \right\rbrack.
 \end{aligned}
\end{equation}
We deal with the Hodge dual by combining
\begin{equation}\label{eq:useful_Hodge_identity}
  \star (\alpha_p \wedge \omega_q) = \frac{1}{p!} \alpha^{\lambda_1
    \dots \lambda_p} \star \omega_{\kappa_1 \dots \kappa_{d-p-q}
    \lambda_1 \dots \lambda_p} dx^{\kappa_1} \otimes \dots \otimes dx^{\kappa_{d-p-q}}
\end{equation}
with \eqref{eq:Hodge-dual_J} and \eqref{eq:Hodge-dual_Reeb}. The
result is
\begin{equation}
  \begin{aligned}
    \imath \star dv_1 &= \imath q v_2 - \imath q v_3 - \frac{q}{4}
    \omega, \\
    \imath \star dv_2 &= -q v_2 -2 v_3, \\
    \imath \star dv_3 &= -(q+2) v_2 - \imath \omega.
  \end{aligned}
\end{equation}
Here we defined
\begin{equation}
  \omega \equiv (\tgd\tgdb f)^{\lambda_1 \lambda_2} J \wedge
    J_{\kappa_1\kappa_2\lambda_1\lambda_2} dx^{\kappa_1} \otimes
    dx^{\kappa_2} = -4 v_1 - \imath (\Delta_0 - 4q - q^2) v_2.
\end{equation}
The second equality here can be derived using the projection operators
\eqref{eq:SE-conventions__projection_operators}. Putting everything
together, one finds
\begin{equation}
  Q v_i = M_{ij} v_j, 
\end{equation}
with
\begin{equation}
  M_{ij} =
  \begin{pmatrix}
    q & \frac{\imath q\lbrack \Delta_0 + 4 - q (q+4) \rbrack}{4} &
    -\imath q \\
    0 & -q & -2 \\
    2 \imath & -\frac{\Delta_0 + 4 - q (2+q)}{2} & 0
  \end{pmatrix}.
\end{equation}
Once again, we diagonalize $M$ and find
\begin{equation}
  \begin{array}{|c|l|}
    \hline
    \text{Mode} & \text{Eigenvalue} \\
    \hline
    fJ^0 & 0 \\
    fJ^- & -(E_0+2) \\
    fJ^+ & E_0+2 \\
    \hline
  \end{array}
\end{equation}
The first of these is a gauge mode, while the others are a bit more complicated:
\begin{equation}
  \begin{aligned}
    fJ^0 &= q ( -2 \imath v_1 - q v_2 + q v_3 ) = - 2 \imath q (d\tgdb f), \\
    fJ^- &= \frac{E_0 + 2 - q}{4} \left\{ - 4 \imath v_1 + \left\lbrack 4 + E_0
    (E_0 + 4 + q) \right\rbrack v_2 + 2 (E_0 + 2 + q) v_3 \right\}, \\
    fJ^+ &= \frac{E_0 + 2 + q}{4} \left\{ 4 \imath v_1 - \left\lbrack (E_0+2)^2
    - (E_0+4)q \right\rbrack v_2 + 2 (E_0 + 2 -q) v_3 \right\}.
  \end{aligned}
\end{equation}
If $f$ is holomorphic, we have $v_2 = v_3$, $v_1 = 0$, and $E_0 =
q$. In this case only $fJ^- = (q+2)^2 v_2$ is non-vanishing.

\subsubsection*{R-charge $r+2$}

We proceed by considering
\begin{equation}
  f \Omega, \quad
  \tgd\tgdb f.\Omega, \quad
  \tgdb(\tgdb f.\Omega), \quad
  \eta \wedge (\tgdb f . \Omega).
\end{equation}
Again, this is not an ideal basis for calculating the action of
$Q$. From \eqref{eq:def_degree_forms_and_projections}, it follows that
\begin{equation}
  \begin{aligned}
    \tgd\tgdb f .\Omega &= \frac{1}{2} \Pi_\mu^{+\kappa} \nabla_\kappa
    \nabla^\lambda f \Omega_{\lambda\nu} dx^\mu \wedge dx^\nu + q f
    \Omega = \frac{1}{2} \tgd ( \tgdb f.\Omega ), \\
    \tgdb(\tgdb f . \Omega) &= \Pi_\mu^{-\kappa}
    \nabla_\kappa \nabla^\lambda f \Omega_{\lambda\nu} dx^\mu \wedge dx^\nu.
  \end{aligned}
\end{equation}
Clearly, $\tgd\tgdb f.\Omega$ is a $(2,0)$ form and it follows that
there is a function $h$ such that
\begin{equation}
  \tgd(\tgdb f.\Omega) = h  f \Omega.
\end{equation}
Contracing with $\bar{\Omega}$ using
\eqref{eq:omega__omega-bar__identity} gives
\begin{equation}
  h = \frac{-\Delta_0 + q^2 + 4q}{2}.
\end{equation}
So $h$ vanishes for holomorphic $f$.

In the end, we choose the following basis:
\begin{equation}
  v_1 = f\Omega, \quad
  v_2 = (\tgd + \tgdb)(\tgdb f.\Omega), \quad
  v_3 = \eta \wedge (\tgdb f. \Omega),
\end{equation}
which allows us to immediately anticipate the form of the gauge mode:
\begin{equation}
  f\Omega^0 \equiv v_2 + \imath (q+3) v_3.
\end{equation}
Also, note that $(\tgd - \tgdb)(\tgdb f.\Omega) = (-\Delta_0 + q^2 + 4q)
v_1 - v_2$. Finally, the following identity is quite useful:
\begin{equation}\label{eq:double_J_identity}
  \begin{aligned}
    &\alpha^{\lambda_1 \lambda_2} J \wedge J_{\kappa_1 \kappa_2
      \lambda_1 \lambda_2} dx^{\kappa_1} \otimes dx^{\kappa_2} \\
    &= 2 J \alpha^{\mu\nu} J_{\mu\nu} \\
    &+ 4 \left(
      \frac{\Pi_{\kappa_1}^{+\lambda_1}
        \Pi_{\kappa_2}^{+\lambda_2}}{2} + \frac{\Pi_{\kappa_1}^{-\lambda_1}
        \Pi_{\kappa_2}^{-\lambda_2}}{2} - \Pi_{\kappa_1}^{+\lambda_1}
      \Pi_{\kappa_2}^{-\lambda_2} \right) \alpha_{\lbrack \lambda_1
      \lambda_2 \rbrack} dx^{\kappa_1} \wedge dx^{\kappa_2} \\
    &= 2 J \alpha^{\mu\nu} J_{\mu\nu} + 4 \left( \alpha^{(2,0)} +
      \alpha^{(0,2)} - \alpha^{(1,1)} \right),
  \end{aligned}
\end{equation}
where the first equality holds for generic tensors $\alpha$, while the
second concerns only two-forms.

Direct calculation yields
\begin{equation}
  \begin{aligned}
    dv_1 &= df \wedge \Omega + 3 \imath f \eta \wedge \Omega, \\
    dv_2 &= -2\imath (q+3) J \wedge (df.\Omega) + \imath (q+3) \eta
    \wedge (\tgd+\tgdb)(df.\Omega), \\
    dv_3 &= 2 J \wedge (\tgdb f.\Omega) - \eta\wedge d(\tgdb f.\Omega),
  \end{aligned}
\end{equation}
and
\begin{equation}
  \begin{aligned}
    \imath \star d v_1 &= - (q+3) v_1 - \imath v_3, \\
    \imath \star d v_2 &= -(q+3) (-\Delta_0 + q^2 + 4q) v_1 + (q+3)
    v_2 + 2\imath (q+3) v_3, \\
    \imath \star d v_3 &= -\imath (-\Delta_0 + q^2 + 4q) v_1 + \imath
    v_2 - 2 v_3.
  \end{aligned}
\end{equation}

This time, the matrix $M$ ($Q v_i = M_{ij} v_j$) is given by
\begin{equation}
   M_{ij} =
    \begin{pmatrix}
      -(q+3) & 0 & - \imath \\
      -(q+3)(-\Delta_0 + q^2 + 4q) & q+3 & 2\imath (q+3) \\
      -\imath (-\Delta_0 + q^2 + 4q) & \imath & -2
    \end{pmatrix}
\end{equation}
and diagonalization leads to
\begin{equation}
  \begin{array}{|c|l|}
    \hline
    \text{Mode} & \text{Eigenvalue} \\
    \hline
    f\Omega^0 &  0 \\
    f\Omega^- &  -(E_0 + 3) \\
    f\Omega^+ &  E_0 + 1 \\
    \hline
  \end{array}
\end{equation}
$f\Omega^0$ is of course the gauge mode. As to the others,
\begin{equation}
  \begin{aligned}
    f\Omega^- &= (E_0 - q) \left\lbrack (E_0 + 3) (E_0 + q +4) v_1 +
      v_2 + \imath (E_0 + q + 6) v_3 \right\rbrack, \\
    f\Omega^+ &= v_2 + 2 \imath v_3 + (E_0 - q) \left\lbrack (E_0 + 1) v_1 - \imath
      v_3 \right\rbrack.
  \end{aligned}
\end{equation}
In the holomorphic case, $v_2$
and $v_3$ disappear, as do $f\Omega^-$ and the function $h$.

The eigenmodes with R-charge $r-2$ can be calculated by considering
$\imath \star d (\bar f \Omega)$ etc.~and taking the complex
conjugate. Note that this procedure gives the negative of the actual
values, since $\imath\star d$ changes its sign under complex conjugation.

\section{Cyclic homology of  the Calabi-Yau } \label{abstractmath}
Here, we continue the last paragraph of section \ref{ginzburg}, and (attempt to) rewrite  the single trace index of the quiver $Q$ with the potential $W$ in terms of geometric quantities on the Calabi-Yau cone $X$. 
Admittedly there are many mathematical gaps in the argument; we will at least state where the gaps lie. 

In that section, we introduced Ginzburg's  DG algebra $\fD$ constructed from the modified quiver $\hat Q$ with the differential $\delta$ determined by the superpotential $W$.
We now use the fact
\begin{equation}
H_i(\fD_\text{cyc},\delta)=\overline{HC}_i(\fD)
\end{equation}  where
$\overline{HC}_\bullet(\fD)$ is the reduced cyclic homology of a DG algebra, see \cite{MR923137}. 
This equality holds because $\fD$ is a free commutative DG algebra. 
The non-reduced homology $HC(A)$ of an algebra over $\bC$ satisfies \begin{equation}
HC_{2n-1}(A)=\overline{HC}_{2n-1}(A), \qquad HC_{2n}(A)=\bC\oplus \overline{HC}_{2n}(A),
\end{equation} where $n$ is a positive integer.
We also need to use Hochschild homology $HH_{\bullet}$ below.

From the quiver $Q$ with superpotential $W$, we can define another algebra
\begin{equation}
\fA=\bC Q/dW, 
\end{equation} which is the path algebra generated by the monomials $x_e$ associated to the edge of the quiver $Q$, modulo the F-term relations coming from the derivative of the superpotential. 
When the pair $(Q,W)$ describes the Calabi-Yau cone $X$, the algebra $\fA$ satisfies a mathematical condition called \emph{3-Calabi-Yau}, which in particular implies \cite{Ginzburg:2006fu}  \begin{equation}
\overline{HC}_i(\fD)=\overline{HC}_i(\fA)=\overline{HC}_i(Z),\ \text{and}\ HH_i(\fD)=HH_i(\fA)=HH_i(Z).
\end{equation}
Here, $Z$ is the crepant resolution of the Calabi-Yau cone $X\cup \{0\}$ where $0$ is the tip.

We need to use the long-exact sequence $\cdots\to HC_{n+2}\to HC_n\to HH_n\to HH_{n-1}\to\cdots$  relating the cyclic homology and the Hochschild homology. In our case, we know  \cite{Ginzburg:2006fu,MR2330156} that $\overline{HC}_{\bullet\ge 3}(\fD)=0$, $HH_{\bullet\ge 4}(\fD)=0$ and $\overline{HC}_2(\fD)=HH_3(\fD)$. We also have $HC_0(\fD)=HH_0(\fD)$.
The remaining relevant part of the long exact sequence is \begin{equation}\begin{aligned}
0\to HC_1(\fD)\to HH_2(\fD)\to HC_2(\fD) &\stackrel{S}{\to} HC_0(\fD) \\
&\to HH_1(\fD) \to HC_1(\fD) \to 0.
\end{aligned}\end{equation} Goodwillie's theorem implies the map $S$ is the zero map, so the long exact sequence splits into short exact sequences \cite{MR793184, MR1269324}. 
Then we have \begin{align}
HH_0&=HC_0, \\
HH_1&=HC_0\oplus HC_1,\\
HH_2&=HC_1\oplus HC_2,\\
HH_3&=\overline{HC}_2
\end{align} in our case.
We also have the Hodge decomposition of the  Hochschild homology \begin{equation}
HH_i(Z)=\oplus_{q-p=i} H^q(Z,\wedge^p \Omega_Z).
\end{equation}

Now, we \emph{assume} the natural map $H^q(Z,\wedge^p \Omega_Z)\to H^q(X,\wedge^p \Omega_X)$ is an isomorphism for $0\le p-q\le 3$. (Again, the AdS/CFT correspondence fails if this is not the case.)
On the cone $X$ there is a globaly-defined holomorphic vector field $D$, using which we can globally split \begin{equation}
\Omega_X=\Omega'_X\oplus O_X.
\end{equation} 
Combining these isomorphisms, we have  \begin{equation}
HC_{i}(\fD)=\oplus_{p-q=i} H^q(X,\wedge^p \Omega'_X).
\end{equation}

\end{appendix}

\bibliographystyle{my-h-elsevier}

\begin{thebibliography}{10}





\bibitem{Romelsberger:2005eg}
 C.~Romelsberger,
 Counting chiral primaries in N = 1, d=4 superconformal field theories,
 Nucl.\ Phys.\  B {\bf 747}, 329 (2006)
 [arXiv:hep-th/0510060].

\bibitem{Kinney:2005ej}
 J.~Kinney, J.~M.~Maldacena, S.~Minwalla and S.~Raju,
 An Index for 4 dimensional super conformal theories,
 Commun.\ Math.\ Phys.\  {\bf 275}, 209 (2007)
 [arXiv:hep-th/0510251].

\bibitem{Spiridonov:2010em}
 V.~P.~Spiridonov,
 Elliptic beta integrals and solvable models of statistical mechanics,
 Contemp.\ Math.\  {\bf 563}, 181 (2012)
 [arXiv:1011.3798 [hep-th]].

\bibitem{Festuccia:2011ws}
 G.~Festuccia and N.~Seiberg,
 Rigid Supersymmetric Theories in Curved Superspace,
 JHEP {\bf 1106}, 114 (2011)
 [arXiv:1105.0689 [hep-th]].

\bibitem{Romelsberger:2007ec}
 C.~Romelsberger,
 Calculating the Superconformal Index and Seiberg Duality,
 arXiv:0707.3702 [hep-th].

\bibitem{Dolan:2008qi}
 F.~A.~Dolan and H.~Osborn,
 Applications of the Superconformal Index for Protected Operators and
 q-Hypergeometric Identities to N=1 Dual Theories,
 Nucl.\ Phys.\  B {\bf 818}, 137 (2009)
 [arXiv:0801.4947 [hep-th]].

\bibitem{Spiridonov:2009za}
 V.~P.~Spiridonov and G.~S.~Vartanov,
 Elliptic Hypergeometry of Supersymmetric Dualities,
 Commun.\ Math.\ Phys.\  {\bf 304}, 797 (2011)
 [arXiv:0910.5944 [hep-th]].

\bibitem{Spiridonov:2011hf}
 V.~P.~Spiridonov, G.~S.~Vartanov and G.~S.~Vartanov,
 Elliptic hypergeometry of supersymmetric dualities II. Orthogonal groups,
 knots, and vortices,
 arXiv:1107.5788 [hep-th].

\bibitem{Sudano:2011aa}
 M.~Sudano,
 The Romelsberger Index, Berkooz Deconfinement, and Infinite Families of
 Seiberg Duals,
 JHEP {\bf 1205}, 051 (2012)
 [arXiv:1112.2996 [hep-th]].

\bibitem{Spiridonov:2012ww}
 V.~P.~Spiridonov and G.~S.~Vartanov,
 Elliptic hypergeometric integrals and 't Hooft anomaly matching
 conditions,
 JHEP {\bf 1206}, 016 (2012)
 [arXiv:1203.5677 [hep-th]].

\bibitem{Klebanov:1998hh}
 I.~R.~Klebanov and E.~Witten,
 Superconformal field theory on three-branes at a Calabi-Yau singularity,
 Nucl.\ Phys.\  B {\bf 536}, 199 (1998)
 [arXiv:hep-th/9807080].

\bibitem{Gauntlett:2004yd}
 J.~P.~Gauntlett, D.~Martelli, J.~Sparks and D.~Waldram,
 Sasaki-Einstein metrics on S**2 x S**3,
 Adv.\ Theor.\ Math.\ Phys.\  {\bf 8}, 711 (2004)
 [arXiv:hep-th/0403002].

\bibitem{Cvetic:2005ft}
 M.~Cvetic, H.~Lu, D.~N.~Page and C.~N.~Pope,
 New Einstein-Sasaki spaces in five and higher dimensions,
 Phys.\ Rev.\ Lett.\  {\bf 95}, 071101 (2005)
 [arXiv:hep-th/0504225].

\bibitem{Benvenuti:2004dy}
 S.~Benvenuti, S.~Franco, A.~Hanany, D.~Martelli and J.~Sparks,
 An Infinite family of superconformal quiver gauge theories with
 Sasaki-Einstein duals,
 JHEP {\bf 0506}, 064 (2005)
 [arXiv:hep-th/0411264].

\bibitem{Franco:2005sm}
 S.~Franco, A.~Hanany, D.~Martelli, J.~Sparks, D.~Vegh and B.~Wecht,
 Gauge theories from toric geometry and brane tilings,
 JHEP {\bf 0601}, 128 (2006)
 [arXiv:hep-th/0505211].

\bibitem{Martelli:2006yb}
 D.~Martelli, J.~Sparks and S.~T.~Yau,
 Sasaki-Einstein manifolds and volume minimisation,
 Commun.\ Math.\ Phys.\  {\bf 280}, 611 (2008)
 [arXiv:hep-th/0603021].

\bibitem{Nakayama:2005mf}
 Y.~Nakayama,
 Index for orbifold quiver gauge theories,
 Phys.\ Lett.\  B {\bf 636}, 132 (2006)
 [arXiv:hep-th/0512280].

\bibitem{Nakayama:2006ur}
 Y.~Nakayama,
 Index for supergravity on AdS(5) x T**1,1 and conifold gauge theory,
 Nucl.\ Phys.\  B {\bf 755}, 295 (2006)
 [arXiv:hep-th/0602284].

\bibitem{Gadde:2010en}
 A.~Gadde, L.~Rastelli, S.~S.~Razamat and W.~Yan,
 On the Superconformal Index of N=1 IR Fixed Points: A Holographic Check,
 JHEP {\bf 1103}, 041 (2011)
 [arXiv:1011.5278 [hep-th]].

\bibitem{Kim:1985ez}
 H.~J.~Kim, L.~J.~Romans and P.~van Nieuwenhuizen,
 The Mass Spectrum of Chiral N=2 D=10 Supergravity on S**5,
 Phys.\ Rev.\  D {\bf 32}, 389 (1985).

\bibitem{Ceresole:1999zs}
 A.~Ceresole, G.~Dall'Agata, R.~D'Auria and S.~Ferrara,
 Spectrum of type IIB supergravity on AdS(5) x T**11: Predictions on N=1
 SCFT's,
 Phys.\ Rev.\  D {\bf 61}, 066001 (2000)
 [arXiv:hep-th/9905226].

\bibitem{Ceresole:1999ht}
 A.~Ceresole, G.~Dall'Agata and R.~D'Auria,
 K K spectroscopy of type IIB supergravity on AdS(5) x T**11,
 JHEP {\bf 9911}, 009 (1999)
 [arXiv:hep-th/9907216].

\bibitem{Ginzburg:2006fu}
 V.~Ginzburg,
 Calabi-Yau algebras,
 arXiv:math/0612139.

\bibitem{Cachazo:2002ry}
 F.~Cachazo, M.~R.~Douglas, N.~Seiberg and E.~Witten,
 Chiral rings and anomalies in supersymmetric gauge theory,
 JHEP {\bf 0212}, 071 (2002)
 [arXiv:hep-th/0211170].

\bibitem{Dijkgraaf:2002dh}
 R.~Dijkgraaf and C.~Vafa,
 A Perturbative window into nonperturbative physics,
 arXiv:hep-th/0208048.

\bibitem{Cox}
D.~A.~Cox, J.~B.~Little and H.~K.~Schenck,
Toric varieties,
Graduate Series in Mathematics vol.~124, AMS, (2011)

\bibitem{MR495499}
V.~I.~Danilov,
The geometry of toric varieties,
Uspekhi Mat.~Nauk, {\bf 33} (1978) 2(200)


\bibitem{Benvenuti:2005cz}
 S.~Benvenuti and M.~Kruczenski,
 Semiclassical strings in Sasaki-Einstein manifolds and long operators in
 N=1 gauge theories,
 JHEP {\bf 0610}, 051 (2006)
 [arXiv:hep-th/0505046].

\bibitem{Wijnholt:2002qz}
 M.~Wijnholt,
 Large volume perspective on branes at singularities,
 Adv.\ Theor.\ Math.\ Phys.\  {\bf 7}, 1117 (2004)
 [arXiv:hep-th/0212021].

\bibitem{Wijnholt:2005mp}
 M.~Wijnholt,
 Parameter space of quiver gauge theories,
 arXiv:hep-th/0512122.

\bibitem{MR2355031}
Raf Bocklandt,
Graded Calabi Yau algebras of dimension 3,
J.~Pure Appl. Algebra {\bf 212} (2008) 1

\bibitem{Eager:2010yu}
 R.~Eager,
 Equivalence of A-Maximization and Volume Minimization,
 arXiv:1011.1809 [hep-th].

\bibitem{MR847717}
R.~P.~Stanley,
Enumerative combinatorics. {V}ol. {I},
Wadsworth \& Brooks/Cole Advanced Books \& Software (1986)

\bibitem{Berenstein:2000ux}
 D.~Berenstein, V.~Jejjala and R.~G.~Leigh,
 Marginal and relevant deformations of N=4 field theories and noncommutative
 moduli spaces of vacua,
 Nucl.\ Phys.\  B {\bf 589}, 196 (2000)
 [arXiv:hep-th/0005087].

\bibitem{Berenstein:2000te}
 D.~Berenstein, V.~Jejjala and R.~G.~Leigh,
 Noncommutative moduli spaces, dielectric tori and T duality,
 Phys.\ Lett.\  B {\bf 493}, 162 (2000)
 [arXiv:hep-th/0006168].

\bibitem{Berenstein:2001jr}
 D.~Berenstein and R.~G.~Leigh,
 Resolution of stringy singularities by noncommutative algebras,
 JHEP {\bf 0106}, 030 (2001)
 [arXiv:hep-th/0105229].

\bibitem{Berenstein:2002fi}
 D.~Berenstein and M.~R.~Douglas,
 Seiberg duality for quiver gauge theories,
 arXiv:hep-th/0207027.

\bibitem{MR2330156}
P.~Etingof and V.~Ginzburg,
Noncommutative complete intersections and matrix integrals,
Pure Appl. Math. Q., {\bf 3} (2007) 1, Special Issue: In honor of Robert D. MacPherson. Part 3

\bibitem{MR1191088}
K.~Igusa,
Cyclic homology and the determinant of the {C}artan matrix,
J.~Pure Appl.~Algebra, {\bf 83} (1992) 2

\bibitem{vanNieuwenhuizen:2012zk}
 P.~van Nieuwenhuizen,
 Spherical harmonics for the compactification of IIB supergravity on S$_5$,
 arXiv:1206.2667 [hep-th].

\bibitem{Larsson2004}
H.~Larsson,
Kaluza-Klein mass spectra, p-branes and AdS/CFT,
M.~Sc.~Thesis (2004)

\bibitem{MR0177135}
J.~J.~Kohn and H.~Rossi,
On the extension of holomorphic functions from the boundary of a
complex manifold,
Ann.~of Math.~(2), {\bf 81} (1965)

\bibitem{MR604043}
S.~S.~T.~Yau,
Kohn-{R}ossi cohomology and its application to the complex {P}lateau problem. {I},
Ann.~of Math.~(2), {\bf 113} (1981) 1

\bibitem{Pope:1982ad}
 C.~N.~Pope,
 Kahler manifolds and quantum gravity,
 J.\ Phys.\  A {\bf 15}, 2455 (1982).

\bibitem{Kihara:2005nt}
 H.~Kihara, M.~Sakaguchi and Y.~Yasui,
 Scalar Laplacian on Sasaki-Einstein manifolds Y**p,q,
 Phys.\ Lett.\  B {\bf 621}, 288 (2005)
 [arXiv:hep-th/0505259].

\bibitem{Oota:2005mr}
 T.~Oota and Y.~Yasui,
 Toric Sasaki-Einstein manifolds and Heun equations,
 Nucl.\ Phys.\  B {\bf 742}, 275 (2006)
 [arXiv:hep-th/0512124].

\bibitem{Baumann:2010sx}
 D.~Baumann, A.~Dymarsky, S.~Kachru, I.~R.~Klebanov and L.~McAllister,
 D3-brane Potentials from Fluxes in AdS/CFT,
 JHEP {\bf 1006}, 072 (2010)
 [arXiv:1001.5028 [hep-th]].

\bibitem{Yamazaki:2012cp}
 M.~Yamazaki,
 Quivers, YBE and 3-manifolds,
 JHEP {\bf 1205}, 147 (2012)
 [arXiv:1203.5784 [hep-th]].

\bibitem{Terashima:2012cx}
 Y.~Terashima and M.~Yamazaki,
 Emergent 3-manifolds from 4d Superconformal Indices,
 Phys.\ Rev.\ Lett.\  {\bf 109}, 091602 (2012)
 [arXiv:1203.5792 [hep-th]].

\bibitem{Kapustin:2004df}
 A.~Kapustin and L.~Rozansky,
 On the relation between open and closed topological strings,
 Commun.\ Math.\ Phys.\  {\bf 252}, 393 (2004)
 [arXiv:hep-th/0405232].

\bibitem{Gaiotto:2009gz}
 D.~Gaiotto and J.~Maldacena,
 The Gravity duals of N=2 superconformal field theories,
 arXiv:0904.4466 [hep-th].

\bibitem{Benini:2009mz}
 F.~Benini, Y.~Tachikawa and B.~Wecht,
 Sicilian gauge theories and N=1 dualities,
 JHEP {\bf 1001}, 088 (2010)
 [arXiv:0909.1327 [hep-th]].

\bibitem{Bah:2012dg}
 I.~Bah, C.~Beem, N.~Bobev and B.~Wecht,
 Four-Dimensional SCFTs from M5-Branes,
 JHEP {\bf 1206}, 005 (2012)
 [arXiv:1203.0303 [hep-th]].

\bibitem{MR923137}
B.~L.~Fe{\u\i}gin and B.~L.~Tsygan,
Cyclic homology of algebras with quadratic relations, universal
enveloping algebras and group algebras,
{$K$}-theory, arithmetic and geometry ({M}oscow, 1984--1986), Lecture
Notes in Math., Springer, {\bf 1289} (1987)

\bibitem{MR2330156}
P.~Etingof and V.~Ginzburg,
Noncommutative complete intersections and matrix integrals,
Pure Appl.~Math.~Q. {\bf 3} (2007) 1, Special Issue: In honor of Robert D. MacPherson. Part 3

\bibitem{MR793184}
T.~G. Goodwillie, ``Cyclic homology, derivations, and the free loopspace,''
  {{\em Topology}
  {\bfseries 24} no.~2, (1985) 187--215}.

\bibitem{MR1269324}
C.~A. Weibel, {\em An introduction to homological algebra}, vol.~38 of {\em
  Cambridge Studies in Advanced Mathematics}.
\newblock Cambridge University Press, Cambridge, 1994.

\end{thebibliography}

\end{document}